\documentclass{ws-ijmpa}
\usepackage{amsmath}
\usepackage{graphicx}
\def\p{\partial}
\def\s{\sigma}
\def\S{\Sigma}
\def\G{\Gamma}

\def\D{\Delta}

\def\ld{\lambda}
\def\Ld{\Lambda}
\def\L{\Lambda}

\def\k{\kappa}

\def\Om{\Omega}
\def\ov{\overline}

\def\b{\beta}

\def\a{\alpha}
\def\ast{\alpha^*}

\def\pdellx'{\frac{\partial}{\partial x'}}
\def\pdellw'{\frac{\partial}{\partial w'}}

\newcommand{\be}{\begin{equation}}
\newcommand{\ee}{\end{equation}}
\def\bed{\begin{displaymath}}
\def\eed{\end{displaymath}}
\def\bea{\begin{eqnarray}}
\def\eea{\end{eqncrray}}

\begin{document}


\title{ Confining Quantum Chromodynamics Model for 3-Quark\\ Baryons, New Mass Source, `Proton Spin Crisis' Solution and\\
  Idealized Quark-Lepton Mass Symmetry}

\author{%
 Jong-Ping Hsu  \\
 Department of Physics, University of Massachusetts Dartmouth, \\
 North Dartmouth, MA 02747-2300, USA\\
Leonardo Hsu\\
Department of Chemistry and Physics,  Santa Rosa Junior College,\\
Santa Rosa, CA 95401, USA\\
}

\maketitle

\begin{abstract}
We discuss a model for the relativistic bound states of 3-quark baryons based on confining quantum chromoynamics (QCD) with general Yang-Mills symmetry.  The model postulates that 3-quark states are formed by consecutive 2-body collisions.  For a proton, d and u quarks get together first, and then they capture another u quark so that the d quark is at the core to form a stable proton state with intergral electric charge.  The two u quarks form a quantum spheric shell and move in a confining potential $C(r)= Q' r$ of the core d quark.  The confining potential $C(r)$ is a static solution of new `phase' fields satisfying the fourth-order equation based on general Yang-Mills symmetry.  The two u quarks with the confining linear potentials $C(r)$  in the spherical shell can produce an effective quark Hooke potential $V_{qH}(r)=Qr^2 + V_o$ for the d quark at the core, where Q and $Q'$ are not independent.  The proton mass is assumed to be approximately given by
 $ E(d) + 2E(u)$, which can be obtained analytically from Dirac Hamiltonians involving $V_{qH}(r)$ and $C(r)$ for d and two u quarks respectively.  The model gives a reasonable understanding of roughly 120 baryon masses based on two different coupling constants and one free parameter $V_o$ for sub-spectra specfied by $J^P$.  These results are roughly within 20\% in percent deviation, which appears to be independent of the assumption of color charges.  The confining QCD model also gives the neutron-proton mass difference $\approx 0.6 MeV$.  
   We propose an experimental test of the  proton  structure.  
   \end{abstract}

 



\section{Introduction}
  
We demonstrate that confining 3-quark bound states and baryon mass spectra could be reasonably understood on the basis of general  SU(3) Yang-Mills dynamic symmetry.\cite{1,2}  General Yang-Mills symmetry involves Hamilton's characteristic function with an arbitrary (Lorentz) vector gauge function in gauge transformations.\cite{3,4,5}  It contains the usual gauge symmetries as special cases when the vector gauge functions can be expressed as the space-time derivative of scalar functions.  The confining 3-quark model involves a confining linear potential $C(r)$, which is a static solution\footnote{The static solutions also give a Coulomb-like potential,\cite{1,5} whose effects are small in the present model and can be neglected.} of new `phase' (or gauge) fields with the fourth-order equation based on general Yang-Mills symmetry.\cite{1,5}  

Baryons are usually supposed to be formed in a 3-body collision of quarks.  But, such a 3-quark system is complicated and does not offer any reasonably simple picture to understand the quark bound states.  However, in the present confining quarkdynamics (QD) model, we postulate a new guiding principle that 3-quark states can be formed by consecutive 2-body collisions.  For example, let us consider a proton, denoted by $p^+(d|uu)\equiv p^+(core|shell)$, with a core d quark and two u quarks in the surrounding shell.  One could imagine that d quark and u quark come together first. Before they fly apart, the d quark captures another u quarks, so that they form a stable `quantum spherical shell' with the d quark at the core.  Other consecutive 2-body collisions could form neutrons $n^0(d|ud)$ and $\L(s|ud)$ baryons, etc..  

It has been demonstrated that the two u quarks with the confining potentials in a spherical shell can produce an effective quark Hooke potential 
$
V_{qH}(r)=Qr^2 + V_o  \ 
$
inside, where $V_o$  is a constant of integration.\cite{1,5}  It would affect the motion of the d quark at the core.  For simplicity, the proton energy eigenstate (or mass) is assumed to be approximately given by the sum $ E(d) + 2E(u)$, which can be derived analytically from Dirac Hamiltonians involving $V_{qH}(r)$ and $C(r)$ for d and two u quarks respectively, as we shall see below.  
 
 Based on these ideas, we discussed a model for the N baryon and K meson mass spectra in a previous work.\cite{5}  The coupling constant Q in the quark Hooke potential was assumed to involve a pure phenomenological function $R_{\ell} =\G(\ell/2 +7/2)/\G(\ell/2 + 2)$, which depends on the angular momentum quantum number $\ell$.  In the present simplified and  improved version of the model, we do not assume this additional phenomenological function $R_{\ell}$ in the quark Hooke potential.  This simplification makes the connection between Q and $Q'$ (in confining potential) more clear and definite.  Furthermore, this also helps to reveal a new mass source due to the strong interactions of quarks via the linear potential C(r) and the quark Hooke potential in the quark Dirac Hamiltonians.  
  
 The coupling constant Q in the quark Hooke's potential is not independent of the coupling constant $Q'$ in the confining potential.  Both Q and Q' involve the parameter K in the calculations of the baryon mass spectra.   The model assumes that $V_o$ (or b) should be the same for a specific sub-spectrum, which involves 1 to 6 eigenstates of the baryon masses with the same total angular momentum quantum number J and parity P, $J^P$.  It is gratifying that one parameter $V_o$ and two coupling constants of the  can fit roughly 120 baryon masses (i.e., N baryons, $\D$ baryons and $\L$ baryons, etc. with masses  from $\approx$1000 MeV to $\approx$ 6000 MeV). 
 
The confining QD model reveals an interesting new source of mass in the physical universe.  Almost all observable masses in the universe appear to be carried by the omnipresent protons and neutrons.  According to the model, the proton is made of three light quarks, i.e., one core d quark and two shell u quarks. These three quarks in a proton have a total masses of roughly 10 MeV $(\approx m_d + 2 m_u=[4.67 +2(2.16)] MeV.)$ However, the proton mass is about 1000 MeV.  Where does this additional mass come from?  One knows that the relativistic motion of a particle can increase its mass.  But so far it appears that this relativistic motion does not lead to the proton mass of roughly 1000 MeV and does not give numerically the baryon mass spectrum.  However, as we shall see from the energy eigenvalues of baryons below, the confining QD shows that about 95 \% of the proton mass is due to the strong interactions between the core d quark and the shell u quarks through the quark Hooke potential $V_{qH}=Qr^2 + V_o$ with the large coupling constant $Q\approx 2.5 (10^7) MeV^3$.  In other words, the model gives us a picture that the light d quark (with mass $m_d$= 4.67 MeV) at the proton core interacting with the quark Hooke potential becomes effectively a massive particle, whose mass is  about 190 times larger than $m_d$.   In contrast, the u quark in the surrounding quantum shell will increase its mass to  about 10 times larger than its original mass $m_u=2.16 MeV$.
Consequently, the confining quarkdynamics implies that about 99\% of the observable mass in the universe is originated from the strong quark interactions inside the cosmologically ominipresent of protons and neutrons in stars, etc.

Interestingly, the idea of  quark confinement by linear and quark Hooke potentials was stimulated by a cosmological model of dark energy.\cite{1,5}  Namely,  the linear repulsive Okubo force generated by the omnipresent nucleons with baryonic charges in a galaxy acting on a supernova.  Such a cosmic Okubo force is generated by the conserved baryon number (or charge), which is postulated to be associated with general U(1) Yang-Mills symmetry.\cite{2}  Such a repulsive linear OKubo force with extremely small baryon charge, which cannot be detected in laboratory, can overcome the gravitational attractive force when the distance separation and the number of nucleons involved became sufficiently large.  Thus, the cosmic Okubo force can lead to the observable effects in the motion of galaxies, i.e., the effects of `dark energy', or the late-time accelerated expansion of the observable `matter half-universe' in the HHK model of Big-Jets for the beginning of the universe. \cite{1,5}

 \section{Relativistic quark Hamiltonian with a confining linear potential}
   The confining QD model involves (a) the quark Hooke potential $V_{qH}$ with a constant of integration $V_o$,\cite{5}   and (b) the confining linear potential $C(r)$,
\be
V_{qH}(r) \equiv Q r^2 + V_o, \ \ \ \   c=\hbar =1,
 \ee
 $$
 Q=\frac{16g^2_s \sqrt{Q'}}{18 L_s^2}\approx  20.28 K \times 10^8 MeV^3, \ 
 $$
 \be
  C(r)=Q'r,
 \ee
$$
 Q'=\frac{g_s^2 K^2}{8\pi L_s^2}\approx 20.2 K^2\times 10^4 MeV^2,   
 $$
$$
1/fm\approx197 MeV, \ \ \  L_s \approx 0.082 fm, \ \ \ \ \  g^2_s/(4\pi)\approx 0.07.
$$
 The constant $V_o$ is important for the spacings of energy eigenvalues of a relativistic Hamiltonian.  We introduce a parameter $K^2$ in (2), which may be used to compare the coupling constants in the model with those empirical values obtained by the Cornell group.\cite{5,6}. If the value of K is approximately 1, the the coupling constant for the linear confining quark potential is roughly the same as the emperical potential obtained by the Cornell group  based on the charmonium data.  As we shall see later. the 3-quark confining model gives a much smaller value for $K \approx 0.01$ based on roughly 120 baryon masses.




Consider the proton composed of two u-quarks and one d quark. The confining QD model for 3-quark  baryons postulates that there are two separate interactions, which may be pictured as follows:   

(i) Each of the two u-quarks moving around the core d quark in a confining linear potential $C(r)$ produced by the d quarks.

(ii) The d-quark moving in an effective quark Hooke potential (1), which is produced by the two u-quarks in the quantum shell with the confining potential acting on the d-quark.\cite{1,5}  The quark charge of the two u quarks is, for simplicity, assumed to have the quark charge density $ |\Psi_u|^2 \propto exp(-a^2 r^2).$ 

Let us first consider the relativistic Hamiltonian\cite{7} $H_u$ of a u-quark and solve for the energy eigenvalue $E_u$ with the help of the Sonine-Laguerre equation.\cite{8}  In the confining QD model, the Hamiltonian $H_u$ with the confining linear potential $C(r)=Q'r$ is postulated to be
\be
H_{u} \approx \a_k p_k + \b m + i \a_k e_k \b C(r),\ \ \ \  
\ee
$$
 C=C(r)=\frac{g_s^2 K^2}{8\pi L_s^2} r\equiv Q'r, \ \ \   m= m_u
 $$
 $$
  p_k=-i\p/\p x^k, \ \ \  e_k=x_k/r, \ \ \ \  k=1,2,3,
$$
 $$
  \a_k=\begin{pmatrix}  0 & \s_k\\ \s_k & 0\end{pmatrix}, \ \ \ \  \beta=\begin{pmatrix}I & 0 \\ 0 & -I\end{pmatrix}, 
 $$
 where $\a_k$ and $\b$ are the usual Dirac matrices.\cite{7}     

  To find the energy eigenvalues of the confining QD model, we write as usual the u quark wave function $q$, which satisfies the equation
\be
H_{u} q = E q, \ \ \ \ \   
\ee
\bigskip

 \ \ \ \ \ \ \ \ \ \ \ \ \ \ \   q=\(\begin{pmatrix} q_A\\q_B\end{pmatrix}\)=\(\begin{pmatrix}g(r)r^{-1}Y_A\\if(r)r^{-1}Y_B\end{pmatrix}\),
 
 \bigskip

\noindent
where $Y_{A} \equiv Y^{j_3}_{j\ell_A}$ and $Y_{B} \equiv Y^{j_3}_{j\ell_B}$ are $r$-independent normalized spin-angular functions (or spinor spherical harmonics).\cite{7} The factor $i$ in $if(r)r^{-1}Y_B$ is to make $f(r)$ and $g(r)$ real for bound state solutions.  We have
 \be
 (E - m) \frac{g(r)}{r}Y_A \approx (\s_k p_k-i\s_k e_k C ) \frac{if(r)}{r}Y_B,
 \ee
 $$
 (E  + m)\frac{ if(r)}{r}Y_B \approx (\s_k p_k +i\s_k e_k C) \frac{g(r)}{r} Y_A.
 $$
As usual, we use the phase convention of Condon and Shortly,\cite{7} so that we have
$( \s_k r_k/ r)Y_A = - Y_B$  and $( \s_k r_k/ r)Y_B = - Y_A.$   Moreover, we also have the usual relations\cite{7}
 \be
 (\s_k p_k) q_B= i\frac{\s_k x_k}{r^2}\left( -ir\frac{\p}{\p r}+i \s_k L_k\right)\frac{f}{r} Y_B
 \ee
 $$
 \ \ \ \ \ \ \ \ \ \ \ \ \ \  =-\frac{d(f/r)}{dr} Y_A -\frac{(1-\kappa)}{r^2} f(r) Y_A, \
 $$
   \be
 (\s_k p_k) q_A= i\frac{d(g/r)}{dr} Y_B + i\frac{(1+\kappa)}{r^2} g(r) Y_B,
 \ee
 $$
 \k=(j+1/2)=\ell >0, \ \ \ \ \  \k=-(j+1/2) = -(\ell+1) <0,
 $$
where $L_k$ is the angular momentum operator.   It follows from (5)-(7) that
 \be
  \frac{df}{dr} -\frac{\kappa}{r}f+Cf \approx -(E - m ) g, \ \ 
  \ee
 \be
  \frac{dg}{dr} +\frac{\kappa}{r} g -Cg \approx (E+m) f.
 \ee
in spherical coordinates.  
The conserved spin-orbital coupling quantum number $\kappa$ is a non-zero integer which can be positive or negative.  Roughly speaking, the sign of $\k$ determines whether the spin is antiparallel ($\k>0$) or parallel ($\k <0$) to the total angular momentum in the nonrelativistic limit.\cite{7}  The total angular momentum quantum number $j$ and the angular momentum quantum number $\ell=\ell_A$ of the upper component $q_A$ and the parity $(-1)^{\ell}$ are determined by $\k$.

\section{Basic quark equation with C(r) and energy eigenvalues}
We obtain the r-dependent upper component $g=g(r)$ by eliminating the lower component $f=f(r)$ from equation (8) by using (9) and $C=Q'r$. This yields a basic equation for the u quark in the confining QD model, 
   \be
 \left[\frac{d^2}{dr^2} - \frac{\kappa^2+\kappa}{r^2}+(2\k-1) Q' - Q'^2 r^2 \right.
   \ee
   $$
  \left.  +(E^2 -m^2) \frac{}{}\right] g(r) \approx 0.
   $$
 To find the energy eigenvalues of the u-quark,  we define a new dimensionless variable $y$, which is related to $r^2$ by 
\be
Q' r^2 =y.
\ee
Based on the relations between $\k$ and $\ell$ in (7), we have $\k^2 +\k=\ell^2 +\ell$ for both  $\k=(j+1/2)=\ell >0, $ and $\k=-(j+1/2) = -(\ell+1) <0$.  Thus, equation (10) can be written as   
  \be
   \left[y\frac{d^2}{dy^2} +\frac{1}{2}\frac{d}{dy} - \frac{\ell^2+\ell}{4y}- \frac{y}{4} +\frac{(2\k -1)}{4}\right.
   \ee
   $$
  \left. +\frac{E^2 - m^2 }{4Q'}\right]g(y)\approx 0.
  $$
 We look for a solution to this equation of the form\cite{7}   
  \be
  g(y)= e^{-y/2} y^{[(\ell +1)/2]} G(y),
  \ee
which is consistent with the asymptotic property,  $g(y)\to 0$ as $y\to \infty $. 

 We obtain the following Sonine-Laguerre differential equation\cite{8} for $G(y)$,
  \be
 \left[ y\frac{d^2 }{dy^2} + \left(\ell +\frac{3}{2}-y\right)\frac{d}{dy} - \frac{\ell}{2} - \frac{3}{4} \right.
 \ee
 $$
\left. +\frac{(2\k - 1)}{4} +\frac{E^2-m^2 }{4Q'}\right]G(y) \approx 0.
 $$
  The solutions to the equation (14) are the Sonine-Laguerre polynomials of degree $n$.  Note that 
   n comes from the Sonine-Laguerre equation $yd^2G/dy^2 + (A-y)dG/dy +nG=0$, where n is an integer greater than or equal to zero.   Thus, we have 
  \be
  n = -\left(\frac{\ell}{2} + \frac{3}{4}- \frac{2\k-1}{4} -\frac{E^2-m^2}{4Q'} \right), \ \ \  
  \ee
  $$
  E=E_u, \ \ m=m_u,
  $$
where $n=0,1,2,....$.  The confining QD model gives the energy eigenvalues of one u-quark $E_u=E$ in terms of the principal quantum number $n$, orbital angular momentum quantum number  $\ell$, and a `spin-orbit' coupling quantum number $\k$,   
  \be
  E_u= \left[4Q'\left(n+\frac{\ell-\k}{2} + 1 \right) +m_u^2\right]^{0.5}.
  \ee

\section{Relativistic Hamiltonian with quark Hooke potential and energy eigenvalues of the core quark}
Apart from the energies of two u quarks, the motion of the core d quark in the quark Hooke potential $V_{qH}$, produced by the two u quarks,\cite{1,5} also contributes to the proton mass.  The confining QD model postulates the following Hamiltonian 
$H_d$ for the d-quark,   
 \be 
 H_{d} \approx \a_k p_k + \b m_d + \frac{(1+\b)}{2}  V_{qH},   \ \ \ \    V_{qH}= Q r^2 + V_o.
\ee

To obtain the energy eigenvalues  in (17) of the d-quark, we use $H_d$ in (17) and following the steps from (3)  to (16).  Instead of equations (8), (9) and (10), we have the following equations for the d-quark,
 \be
  \frac{df}{dr} -\frac{\kappa}{r}f \approx -(E - m - V_{qH}) g, \ \ 
  \ee
 \be
  \frac{dg}{dr} +\frac{\kappa}{r} g \approx (E+m) f.
 \ee
   \be
 \left[\frac{d^2}{dr^2} - \frac{\kappa^2+\kappa}{r^2}-(E + m) V_{qH}  \right.
\ee
$$ 
\left. +(E^2 -m^2) \frac{}{}\right] g(r) \approx 0, \ \ \  m=m_d.
$$
Since $ V_{qH}=Q r^2 + V_o$, it is convenient to define a new dimensionless variable $y=\sqrt{[(E+m)Q]} \ r^2\equiv a^2 r^2$, equation (20) can then be written as
  \be
   \left[4y\frac{d^2}{dy^2} + 2\frac{d}{dy} - \frac{\ell^2+\ell}{y}- y \right.
   \ee
   $$
  \ \ \ \ \ \ \ \ \ \   \left. +\frac{E^2 - m^2 -(E+m)V_o}{a^2}\right]g(y)\approx 0.
  $$
To solve (21), we look for a solution of the form, $ g(y)= e^{-y/2} y^{[(\ell +1)/2]} G_1(y)$.  We obtain the Sonine-Laguerre equation\cite{8} for the d-quark located in the core of the poton with the quark Hooke potential  $V_{qH}$,
\be
 \left[ y\frac{d^2 }{dy^2} + \left(\ell +\frac{3}{2}-y\right)\frac{d}{dy} - \frac{\ell}{2} - \frac{3}{4} \right.
 \ee
 $$
 \ \ \ \ \ \ \ \ \   \left.  +\frac{E^2-m^2 -(E+m)V_o}{4a^2}\right]G_1(y) \approx 0, 
$$
$$
a^2=\sqrt{[(E+m)Q]}.
$$
Following steps (12)-(16), we obtain the relativistic equation for the energy eigenvalue $E=E_d, $ of the d quark,
\be
E_d^2 -m_d^2= 4\sqrt{(E_d +m_d)Q}  \ (n+ \frac{\ell}{2} + 0.75+ B),  \ \  m_d =m,
\ee
$$
Q  \approx  20.28 K  \times 10^8 MeV^3,    \ \ \    
$$
$$                             
 B=\frac{(E_d+m_d)V_o}{4a^2} = b \sqrt{E_d+ m_d}, \ \ \ \ \  V_o=4b \sqrt{Q},      
$$
where $V_o$ is expressed in terms of the parameter $b$ and the coupling constant Q for convenience.

\section{Baryon  mass spectra}
There were 28 N baryon energy states listed in the particle data.\cite{10}  The existence of 22 of them is deemed very likely or certain, and in general, their properties are fairly well-explored. Much less is known about the other 7 states, and their existence is much less certain, according to the particle data group P. A. Zyla et al. \cite{10,11}. We list all data for completeness and for future comparisons of the model predictions and possible new data.

Based on (16) and (23), the confining QD model approximates the energy eigenstate $E_n(N)$ of an N baryon to be the sum of the d-quark energy $E_d$ and the two u-quark  energy $E_{2u}$,
 \be
 E_n(N)\approx E_d + E_{2u},
 \ee
 $$
 E_d= \sqrt{4\sqrt{(E_d +m_d)Q}  \ (n+ \frac{\ell}{2} + 0.75+ b\sqrt{E_d+ m_d}) + m^2_d}, 
  $$
  $$
 E_{2u}= 2E_{u}= 2\sqrt{\left[4Q' \left(n+\frac{\ell-\k}{2} + 1 \right) +m_u^2\right]}, \ \ 
 $$
  where
  \be
 Q=(20.28)(K)(10^8)MeV^3, \  \ \      m_d=4.67MeV,
\ee
 $$
    Q'= (20.2)(K^2)(10^4)MeV^2, \ \   m_u=2.16MeV.   
 $$
 The parameter K in the coupling constants Q and $Q'$ are determined by the baryon data in (28)-(90) below.  There are roughly 120 masses in baryon mass spectra.
  
 Suppose we consider a neutron $n^0$ in the model, there are two possible structures, i.e.,  $(u|dd)$ with u quark as the core and $(d|du)$ with d quark as the core.  The core quark is near the center of the effective quark Hooke potential $V_{qH}= Q r^2 + V_o$.  Note that the effective mass of the u-quark in the quark Hooke potential is roughly 931 MeV, which is about 100 times  larger than the two d-quarks with mass $(2 \times 4.67)$MeV in the quantum shell of a neutron.  Thus, the confining QD model indicates that the large  neutron mass is due to the large coupling constant  Q in the quark Hooke potential. 
     
   To be specific, instead of $E_{2u}$ and $E_d$ in (24) for proton, we use the following energy eigenvalue $E(n^0)$ for  a neutron $n^0(u|dd)$,
\be
 E(n^0)\approx E_u + E_{2d},
 \ee
  $$
 (E_{2d})^2= 4\left[4Q'\left(n+\frac{\ell-\k}{2} + 1 \right) +m_d^2\right], \  \  
 $$
$$
 (E_u)^2= 4\sqrt{(E_u +m_u)Q}  \ (n+ \frac{\ell}{2} + 0.75+ b\sqrt{E_u+m_u}) + m^2_u.
 $$
This will be used to calculate the  neutron-proton mass difference based on the confining QD model.  

 \subsection{$N$ baryons with sub-spectrum specified by total angular momentum J and parity P}

All N baryons have strangeness S=0 and isospin I=1/2.  The confining QD model model leads to the following results with a prediction of energy eigenstate beyond the highest state of a 
 given 'sub-spectrum' specified by $J^P$.  All N baryons in the model have the same coupling constants Q and Q'.  However, each sub-spectrum has a specific value for b (or $V_0$) in the quark Hooke potential. 
 
 For $N^+(d|uu)$ baryon spectrum with given total angular momentum $J$ and parity $P$, $J^P$, such as $N1/2^+$ etc.,\cite{3} the confining QD model gives the result (24), i.e., 
 
 \be
 E_n \approx E_d(n) + E_{2u}(n), 
 \ee
 $$
 E^2_d(n) = 4\sqrt{(E_d(n) +4.67)Q}  \ \left[n+ \frac{\ell}{2} + 0.75 \right.
 $$
 $$
\left. + b\sqrt{E_d(n)+ 4.67} \right] + 4.67^2,
 $$
 $$
 E^2_{2u}(n)= 4 \left[4 Q' \left(n+\frac{\ell-\k}{2} + 1 \right) +2.16^2\right],		 
 $$
  $$
 \k=(j+1/2)=\ell >0, \ \ \ \ \  \k=-(j+1/2) = -(\ell+1) <0,
 $$
 where Q and Q' are given in (25).
 
   In the following discussions, the confining QD model gives the approximate results for approximately 120 baryons in (28)-(90) below with two coupling constants (corresponding to K=0.012 and K=0.006) in (25) and one free parameter b for each sub-spectrum.  The results  display reasonably consistent with data within roughly 20\% of percent deviation, i.e., 
   $ |$(data value - model value )/(data value)$|$. 
   
\bigskip  

 In CQD (confining quarkdynamics), quark masses (in MeV) are given by the particle data,\cite{3}
 \bigskip
   
{\bf u(2.16), c(1270), t(172760),
 
 d(4.67),  \ \  s(93),   \ \  b(4180).}
 \bigskip
 
 \noindent
 With the help of WalframAlpha, the CQD results for baryon spectra with roughly 120 masses are calculated and listed below:
     \be
  model:  \ell=0, \   \kappa=-1;   \   p, N^+ =(d|uu),  \ J^P=1/2^+,
  \ee
      $$
    CQD:  K=0.012,\  b= \frac{ 0.02}{\sqrt{MeV}},   \ \ \ \    
   $$
$$
     CQD(MeV),  \ \ \ \ \ \ \ \ \ \ \ \ \ \ \ \ \ \     {\bf data},  
     $$
  $$
    n=0, \ \ \  E_{0}\approx 920MeV, \ \ \ \ \ \ \ \ \ \    {\bf  N(938)},
  $$
  $$
   n=1,   \ \ \ E_{1}  \approx 1375 MeV, \ \ \ \ \ \ \ \ \  {\bf N(1440)}, 
  $$
  $$
     n=2,   \ \ \ E_{2} \approx 1750 MeV, \ \ \ \ \  \ \ \ \ \  {\bf N(1710)}, 
  $$
  $$
 n=3,   \ \ \ E_{3} \approx 2082 MeV, \ \ \ \ \ \ \ \ \ \ \ \   {\bf N(1880)}, 
 $$
  $$
    n=4,   \ \ \ E_{4} \approx 2387 MeV, \ \ \ \ \  \ \ \ \ \ \   {\bf N(2100)}, 
  $$
  $$
     n=5,    \ \ \ E_{5} \approx 2671 MeV, \ \ \ \ \  \ \ \ \ \ \ \   {\bf  N(2300) }, 
  $$
    \bigskip
      \be
   model:   \  \ell=1,  \   \kappa= 1;   \      p, N^+ = (d|uu), \  J^P=1/2^-
  \ee 
    $$
    CQD:  K=0.012, \ b= \frac{ 0.038}{\sqrt{MeV}},  
   $$
$$
    CQD(MeV),  \ \ \ \ \ \ \ \ \ \ \ \ \ \ \ \ \ \    {\bf data},  
      $$
   $$
    n=0, \ \  E_{0}\approx 1432 MeV, \ \ \ \ \ \ \ \ \ \ \ \ \ \  {\bf N(1535)}, 
  $$
  $$
   n=1,   \ \ \ E_{1}   \approx 1830 MeV, \ \ \ \ \ \ \ \ \ \ \ \ \  {\bf N(1650)}, 
  $$
  $$
    n=2,   \ \ \ E_{2} \approx 2177 MeV, \ \ \ \ \  \ \ \ \ \ \ \ \ {\bf N(1895)}, 
  $$
   \bigskip
    \be
   model: \   \ell=1,  \  \kappa= -2; \      p, N^+ =(d|uu), \  J^P=3/2^- 
  \ee 
    $$
    CQD:  K=0.012,  \ b= \frac{ 0.03}{\sqrt{MeV}},   \ \ \ \      \ \ \ \ \ \ \ \ \   
   $$
$$
    CQD(MeV),  \ \ \ \ \ \ \ \ \ \ \ \ \ \ \ \ \   {\bf data},   
      $$
  $$ 
   n=0, \ \  E_{0}\approx 1318 MeV, \ \ \ \ \ \ \ \ \ \ \ \   {\bf  N(1520)},
  $$
  $$
   n=1,   \ \ \ E_{1} \approx 1719 MeV, \ \ \ \ \ \ \ \ \ \ \   {\bf N(1700)},
  $$
  $$
    n=2,   \ \ \ E_{2} \approx 2067 MeV, \ \ \ \ \  \ \ \ \ \ \ \   {\bf N(1875)}, 
  $$
  $$
    n=3,   \ \ \ E_{3} \approx 2382 MeV,  \ \ \ \ \  \ \ \ \ \ \ \ \  {\bf N(2120)},
  $$
      \bigskip
  \be
  model: \ \ell=2, \kappa=2; \   p, N^+ = (d|uu), \  J^P=3/2^+
  \ee
   $$
    CQD:  K=0.012, \ b= \frac{ 0.0285}{\sqrt{MeV}}, 
   $$
$$
    CQD(MeV),  \ \ \ \ \ \ \ \ \ \ \ \  \ \ \ \ \ \ \  {\bf data}, 
      $$
  $$ 
   n=0, \ \  E_{0}\approx 1489 MeV, \ \ \ \ \ \ \ \ \ \ \ \   {\bf  N(1720)},
  $$
  $$
   n=1,   \ \ \ E_{1} \approx 1863 MeV, \ \ \ \ \ \ \ \ \ \ \   {\bf N(1900)}, 
  $$
  $$
    n=2,   \ \ \ E_{2} \approx 2195 MeV, \ \ \ \ \  \ \ \ \ \ \ \   {\bf N(2040)},
  $$  
    $$
   n=3,,   \ \ \ E_{3} \approx 2499 MeV, (prediction).
  $$
  \bigskip
       \be
   model:   \ell=3,    \kappa= 3; \   p, N^+ = (d|uu), \  J^P=5/2^-  
   \ee
    $$
    CQD:  K=0.012, \ b= \frac{ 0.0285}{\sqrt{MeV}},   \ \ \      \ \ \ \ \ \ \ \ \  
      $$
      $$
     CQD(MeV), \ \ \ \ \ \ \ \ \ \ \ \ \ \ \ \ \ \ \ \ {\bf data}, \ 
      $$
   $$
      n=0, \ \  E_{0}\approx 1678 MeV, \ \ \ \ \ \ \ \ \ \ \  {\bf  N(1675)},
  $$
  $$
   n=1,   \ \ \ E_{1}  \approx 2030 MeV, \ \ \ \ \ \ \ \ \     {\bf N(2060)}, 
  $$
  $$
    n=2,   \ \ \ E_{2} \approx 2347 MeV, \ \ \ \ \  \ \ \ \   {\bf N(2570)}, 
  $$
      \bigskip
    \be
   model: \   \ell=2,  \  \kappa= -3; \     p, N^+ = (d|uu), \  J^P=5/2^+
  \ee 
    $$
    CQD:  K=0.012,  \b= \frac{ 0.0285}{\sqrt{MeV}},   \ \ \ \      \ \ \ \ \ \ \ \ \  
   $$
$$
     CQD(MeV),  \ \ \ \ \ \ \ \ \ \ \ \ \ \ \ \  \ \ \ \ \ \ \ \ \ \  {\bf data},  
      $$
   $$
   \ \ n=0, \ \  E_{0}\approx 1508 MeV, \ \ \ \ \ \ \ \ \ \ \ \ \ \  {\bf N(1680)},
  $$
  $$
 \ \ \ \ \  n=1,   \ \ \ E_{1}  \approx 1879 MeV, \ \ \ \ \ \ \ \ \ \ \ \ \  {\bf N(1860)},
  $$
  $$
 \ \ \ \ \   n=2,   \ \ \ E_{2} \approx 2209 MeV, \ \ \ \ \  \ \ \ \ \ \ \ \ {\bf N(2000)},
  $$
      \bigskip    
  \be
   model:   \  \ell=4, \    \kappa= 4; \    p, N^+ = (d|uu), \  J^P=7/2^+
 \ee  
   $$
    CQD:  K=0.012, \ b= \frac{ 0.035}{\sqrt{MeV}},   \ \ \ \     \ \ \ \ \ \ \ \ \   
   $$
$$
    CQD(MeV),  \ \ \ \ \ \ \ \ \ \ \ \ \ \ \   {\bf data},  
      $$
   $$
  n=0, \ \  E_{0}\approx 1950 MeV, \ \ \ \  {\bf N(1990)},
  $$
  $$
   n=1,   \ \ \ E_{1}  \approx 2283 MeV, (prdiction). 
  $$
     \bigskip
      \be
   model:  \   \ell=3, \  \kappa= -4; \     p, N^+ = (d|uu), \  J^P=7/2^-
   \ee
     $$
    CQD:  K=0.012, \ b= \frac{ 0.06}{\sqrt{MeV}},   \ \ \ \ \      \ \ \ \ \ \ \ \ \ 
   $$
$$
    CQD(MeV),  \ \ \ \ \ \ \ \ \ \ \ \ \ \ \ \ \ \   {\bf data},  
      $$
   $$
   n=0, \ \  E_{0}\approx 2191 MeV, \ \ \ \ \ \ \ \ \ \  {\bf N(2190)},
  $$
  $$
    n=1,   \ \ \ E_{1}   \approx 2527 MeV, (prediction), 
  $$
       \bigskip
      \be
   model:  \  \ell=4, \   \kappa= -5; \   p, N^+ = (d|uu), \   J^P=9/2^+
      \ee
     $$
    CQD:  K=0.012, \ b= \frac{ 0.05}{\sqrt{MeV}},   \ \ \ \      \ \ \ \ \ \ \ \ \   
   $$
$$
     CQD(MeV),  \ \ \ \ \ \ \ \ \ \ \ \ \ \ \ \ \ \ \   {\bf data},   
      $$
   $$
   n=0, \ \  E_{0} \approx 2209 MeV, \ \ \ \ \ \ \ \ \  {\bf N(2220)}, 
  $$
  $$
  n=1,   \ \ \ E_{1}   \approx 2532 MeV,  (prediction). 
  $$
      \bigskip
       \be
   model:  \  \ell=5,  \  \kappa= 5; \    p, N^+ = (d|uu), \  J^P=9/2^-
   \ee
     $$
    CQD:  K=0.012, \ b= \frac{ 0.038}{\sqrt{MeV}},   \ \ \ \    \ \ \ \ \ \ \ \ \  
   $$
$$
    CQD(MeV),  \ \ \ \ \ \ \ \ \ \ \ \ \ \ \ \ \ \ \ \ \ \ \   {\bf data},  
      $$
   $$
    n=0, \ \  E_{0})\approx 2206 MeV, \ \ \ \ \ \ \ \ \ \  {\bf  N(2250)},
  $$
  $$
    n=1,   \ \ \ E_{1}   \approx 2523 MeV, (prediction).  
  $$
  \bigskip
  \be
   model:  \   \ell=5, \   \kappa= -6; \   p, N^+ = (d|uu), \   J^P=11/2^-
   \ee
       $$
    CQD:  K=0.012,  \ b= \frac{ 0.07}{\sqrt{MeV}},   \ \ \ \ \     \ \ \ \ \ \ \ \ \  
   $$
$$
    CQD(MeV),  \ \ \ \ \ \ \ \ \ \ \ \ \ \ \ \ \ \ \ \   {\bf data},  
      $$
   $$
   n=0, \ \  E_{0}\approx 2689 MeV, \ \ \ \ \ \ \ \ \ \  {\bf N(2600)},
  $$
  $$
   n=1,   \ \ \ E_{1}   \approx 2992 MeV, (prediction).
  $$
    \bigskip
  \be
   model: \  \ell=6,  \  \kappa= -7; \  p, N^+ = (d|uu), \  J^P=13/2^+
 \ee  
     $$
    CQD:  K=0.012, \ b= \frac{ 0.06}{\sqrt{MeV}},   \ \ \ \      \ \ \ \ \ \ \ \ \  
       $$
$$
    CQD(MeV),  \ \ \ \ \ \ \ \ \ \ \ \ \   {\bf data},  
      $$
   $$
 n=0, \ \  E_{0}\approx 2689 MeV, \ \ \ \  {\bf  N(2700)}, 
  $$
  $$
   n=1,   \ \ \ E_{1}   \approx 2984 MeV, (prediction). 
  $$
  
   
  \subsection{$\D$ baryons }  
 For baryons $  \D^+(d|uu), \  \D^-(d|dd)$ etc. with positive charges and even parity, i.e., $N1/2^+$,the confining QD model gives the following results for $E_n\approx E_d(n) + E_{2u}(n)$ with the corresponding experimental data,
 $$
 E_n\approx E_d(n) + E_{2u}(n),
 $$
 $$
  E^2_d(n)\approx  4\sqrt{(E_d +4.67)Q}  \ \left[ n+ \frac{\ell}{2} \right.
  $$
  $$
  \left. + 0.75+ b\sqrt{E_d+ 4.67}\right] + 4.67^2,
 $$
 $$
 	E^2_{2u}(n)  \approx  4 \left[4 Q' \left(n+\frac{\ell-\k}{2} + 1 \right) +2.16^2\right].	
 $$
 $$
 Q=20.28(K)(10^8)MeV^3, \ \ \ \ \ 
 $$
 $$
  Q'=(20.20)(K^2)(10^4)MeV^2,
 $$
where the contribution of $E_d(n)$ from the d quark at the core is about two orders of magnitude larger than that of $E_{2u}(n)$ contributed from two u quarks in the surrounding spheric quantum cloud.
  \bigskip
    \be
  model:    \ell=1,  \   \kappa= 1; \    \D^+= (d|uu),  \ J^P=1/2^-
  \ee
     $$
    CQD:  K=0.012, \ \  b= \frac{ 0.05}{\sqrt{MeV}},   \ \ \ \ \ \ \  
   $$
$$
     CQD(MeV)     \ \ \ \ \ \ \ \ \ \ \ \ \ \ \ \ \ \ \      {\bf data},
   $$
   $$
   n=0, \ \ \ E_0\approx 1577 MeV, \ \ \ \ \ \ \ \ \    {\bf \D(1620)},
   $$
   $$
   n=1, \ \ \  E_1 \approx 1968 MeV, \ \ \ \ \ \ \ \ \ \   {\bf \D(1900)}, 
   $$
   $$  
     n=2,   \ \ \ E_{2} \approx 2311 MeV, \ \ \ \ \  \ \ \ \    {\bf \D(2150)}, 
  $$
   \bigskip
   \be
   model:   \  \ell=0,  \   \kappa= -1; \    \D^+= (d,uu),  \ J^P=1/2^+
   \ee
     $$
    CQD:  K=0.012,  \ b= \frac{ 0.067}{\sqrt{MeV}},   \ \ \ \ \  
   $$
$$
    CQD(MeV),  \ \ \ \ \ \ \ \ \ \ \ \ \ \ \ \ \ \     {\bf data}, 
      $$
   $$
  n=0, \ \  E_{0} \approx 1713 MeV, \ \ \ \ \ \ \ \ \ \ \     {\bf \D(1750)}, 
  $$
  $$
  n=1,   \ \ \ E_{1}  \approx 2117 MeV, \ \ \ \ \ \ \ \ \ \  {\bf \D(1910)}, 
  $$
  \bigskip
   \be
   model:  \  \ell=1,   \  \kappa= -2; \ \ \ \ \ \      \D^+= (d,uu),  \ J^P=3/2^-
   \ee
     $$
    CQD:  K=0.012,  \ b= \frac{ 0.049}{\sqrt{MeV}},   \ \ \ 
   $$
$$
   CQD(MeV),  \ \ \ \ \ \ \ \ \ \ \ \ \ \ \ \ \ \ \     {\bf data}, 
      $$
   $$
   n=0, \ \  E_{0} \approx 1624 MeV, \ \ \ \ \ \ \ \ \ \ \ \ \    {\bf  \D(1700)}, 
  $$
  $$
   n=1,   \ \ \ E_{1}  \approx 2011 MeV, \ \ \ \ \ \ \ \ \ \    {\bf \D(1940)}, 
  $$
  \bigskip
   \be
   model: \   \ell=2,  \   \kappa= +2; \      \D^+= (d|uu),  \ J^P=3/2^+
   \ee
     $$
    CQD:  K=0.012,  \ b= \frac{ 0.012}{\sqrt{MeV}},   \ \ \ \ \ \  
   $$
$$
    CQD(MeV),  \ \ \ \ \ \ \ \ \ \ \ \ \ \ \ \ \ \   {\bf data},   
      $$
   $$
  n=0, \ \ \  E_{0}\approx 1247 MeV, \ \ \ \ \ \ \ \ \    {\bf  \D(1232)},
  $$
  $$
  n=1,   \ \ \ E_{1}  \approx 1628 MeV, \ \ \ \ \ \ \ \ \ \   {\bf \D(1600)}, 
  $$
  $$
    n=2,   \ \ \ E_{2} \approx 1963 MeV, \ \ \ \ \  \ \ \ \ \ \     {\bf  \D(1920)},
  $$
$$
   n=3,   \ \ \ E_{3} \approx 2269 MeV, (prediction). 
  $$
  \bigskip
   \be
   model:   \     \ell=3,   \  \kappa= 3; \      \D^+= (d|uu),  \ J^P=5/2^- 
   \ee
     $$
    CQD:  K=0.012,  \ b= \frac{ 0.044}{\sqrt{MeV}},   \ \ \ \ \ 
   $$
$$
    CQD(MeV),  \ \ \ \ \ \ \ \ \ \ \ \ \ \ \ \    {\bf data},  
      $$
   $$
   n=0, \ \  E_{0}\approx 1913 MeV, \ \ \ \ \ \ \     {\bf  \D(1930)},
  $$
  $$
  n=1,  \ \  E_{1}  \approx 2260 MeV, \ \ \ \ \ \      {\bf  \D(2350)}, 
  $$
 \bigskip
 \be
   model: \   \ell=2,   \  \kappa= -3; \     \D^+= (d|uu),   \ J^P=5/2^+
  \ee
     $$
    CQD:  K=0.012,  \ b= \frac{ 0.053}{\sqrt{MeV}},   \ \ \ \  
   $$
$$
   CQD(MeV),  \ \ \ \ \ \ \ \ \ \ \ \ \ \ \ \ \ \ \    {\bf data}, 
      $$
   $$
   n=0, \ \  E_{0} \approx 1892 MeV, \ \ \ \  \ \ \ \ \ \        {\bf  \D(1905)},
  $$
  $$
   n=1,   \ \ \ E_{1} \approx 2251 MeV, \ \ \ \ \  \ \ \ \ \ \   {\bf \D(2000)}, 
  $$
   \bigskip
  \be
   model:  \  \ell=3,   \  \kappa= -4; \      \D^+= (d|uu),  \ J^P=7/2^-
   \ee
     $$
    CQD:  K=0.012,  \ b= \frac{ 0.057}{\sqrt{MeV}},   \ \ \ \ \   
   $$
$$
   CQD(MeV),  \ \ \ \ \ \ \ \ \ \ \ \ \ \ \ \ \ \ \   {\bf data},   
      $$
   $$
    n=0, \ \  E_{0} \approx 2144 MeV, \ \ \ \ \ \ \ \ \ \ \  {\bf  \D(2200)}, 
 $$
  $$
   n=1,   \ \ E_{1}  \approx 2480 MeV, (prdiction). \ \ \ \ \ \  \ 
  $$
   \bigskip
  \be
   model:    \  \ell=4,   \  \kappa= 4; \      \D^+= (d|uu),  \ J^P=7/2^+
   \ee
     $$
    CQD:  K=0.012,  \ b= \frac{ 0.04}{\sqrt{MeV}},   \ \ \ \  
   $$
$$
    CQD(MeV),  \ \ \ \ \ \ \ \ \ \ \ \ \ \ \ \ \ \     {\bf data}, 
      $$
   $$
   n=0, \ \  E_{0} \approx 2026 MeV, \ \ \ \ \ \ \ \ \ \   {\bf \D(1950)}, 
  $$
  $$
  \ \ n=1, \ \  E_{1}\approx 2357 MeV, \ \ \ \ \ \ \ \ \ \   {\bf \D(2390)}, 
  $$
  $$
 n=2,   \ \ \ E_{2}   \approx 2660 MeV, (prediction). 
  $$
  \bigskip
   \be
   model:  \  \ell=5,  \   \kappa= 5; \      \D^+= (d|uu),   \ J^P=9/2^-
      \ee
     $$
    CQD:  K=0.012,  \ b= \frac{ 0.055}{\sqrt{MeV}},   \ \ \ \ \  
   $$
$$
    CQD(MeV),  \ \ \ \ \ \ \ \ \ \ \ \ \ \ \ \ \   {\bf data}, 
      $$
   $$
    n=0, \ \  E_{0} \approx 2420 MeV, \ \ \ \ \ \ \ \   {\bf  \D(2400)}, 
  $$
  $$
   n=1,   \ \ \ E_{1}  \approx 2732 MeV,  (prediction). 
   $$
  \bigskip
  \be
   model:   \  \ell=4,   \  \kappa= -5; \      \D^+= (d|uu),  \ J^P=9/2^+
   \ee
     $$
    CQD:  K=0.012,  \ b= \frac{ 0.055}{\sqrt{MeV}},   \ \ \ \ \  
   $$
$$
     CQD(MeV),  \ \ \ \ \ \ \ \ \ \ \ \ \ \ \    {\bf data},  
      $$
   $$
    n=0, \ \  E_{0} \approx 2287 MeV, \ \ \ \ \ \ \    {\bf \D(2300)},
  $$
  $$
   n=1,   \ \ \ E_{1}    \approx 2608 MeV, (prediction). 
  $$
  \bigskip
  \be
   model:  \   \ell=6,  \   \kappa= 6; \     \D^+= (d|uu),  \ J^P=11/2^+
   \ee
     $$
    CQD:  K=0.012,  \ b= \frac{ 0.045}{\sqrt{MeV}},   \ \ \ \ \  
   $$
$$
    CQD(MeV),  \ \ \ \ \ \ \ \ \ \ \ \ \ \ \    {\bf data},  
      $$
   $$
   n=0, \ \  E_{0} \approx 2423 MeV, \ \ \ \ \ \     {\bf  \D(2420)}, 
  $$
  $$
   n=1,   \ \ \ E_{1}   \approx 2727 MeV, (prediction). 
  $$
   \bigskip
  \be
   model:     \ell=7, \   \kappa= 7; \ \      \D^+= (d|uu),  \ J^P=13/2^-
   \ee
     $$
    CQD:  K=0.012,  \ b= \frac{ 0.055}{\sqrt{MeV}},   \ \ \ \ \   
   $$
$$
    CQD(MeV),  \ \ \ \ \ \ \ \ \ \ \ \ \ \    {\bf data},  
      $$
   $$ 
   n=0, \ \  E_{0} \approx 2724 MeV, \ \ \ \ \ \     {\bf  \D(2750)}, 
  $$
  $$
   n=1,   \ \ \ E_{1}  \approx 3015 MeV, (prediction). 
  $$
  \bigskip
   \be
   model:   \   \ell=8,   \  \kappa= 8; \     \D^+= (d|uu),  \ J^P=15/2^+
   \ee
     $$
    CQD:  K=0.012,  \ b= \frac{ 0.06}{\sqrt{MeV}},   \ \ \ \    
   $$
$$
   CQD(MeV),  \ \ \ \ \ \ \ \ \ \ \ \ \ \ \ \     {\bf data},  
      $$
   $$
    n=0, \ \  E_{0} \approx 2943 MeV, \ \ \ \ \ \ \   {\bf  \D(2950)}, 
  $$
  $$
   n=1,   \ \  E_{1}  \approx 3225 MeV, (prediction). 
  $$

\subsection{$\L$ baryons}
 For $\L^{0}(uds)$ baryon states with zero charge and even parity, i.e., $N1/2^+$,the confining QD model gives the following results for $E_n\approx E_s(n) + E_{ud}(n)$ with the corresponding experimental data.  
 \bigskip 
 \be
model:  \ \  \ell=0, \ \ \kappa=-1; \   \L^0=(s|ud), \ J^P=1/2^+
\ee
$$
  CQD : K = 0.006, \ b = \frac{0.054}{\sqrt{MeV}}, \ \ \ \  
  $$
  $$
CQD(MeV ), \ \ \ \ \ \ \ \ \ \ \ \ \   {\bf data}, 
$$
  $$
   n=0, \ \  E_{0} \approx 1157MeV, \ \ \ \ \ \ \ \   {\bf  \L(1115)}, 
  $$
  $$
  n=1,   \ \ \ E_{1}  \approx 1483 MeV, \ \ \ \ \ \ \ \    {\bf \L(1600)}, 
   $$
  $$
   n=2,   \ \ \ E_{2} \approx 1763 MeV, \ \ \ \ \  \ \ \     {\bf \L(1710)}, 
  $$
     $$
   n=3,    \ \ \ E_{3} \approx 2015 MeV, \ \ \ \ \  \ \ \     {\bf \L(1810)}, 
  $$
      $$
   n=4,    \ \ \ E_{4} \approx 2249 MeV, (prediction). \ \ \ \ \  
  $$
  \bigskip 
   \be
   model:   \   \ell=1,   \   \kappa= 1; \   \L^0=(s|ud), \ J^P=1/2^-
   \ee
   $$
  CQD : K = 0.006, \ b = \frac{0.04}{\sqrt{MeV}}, \ \ \ \ 
  $$
  $$
CQD(MeV ), \ \ \ \ \ \ \ \ \ \ \ \ \   {\bf data}, 
$$
  $$
   n=0, \ \  E_{0}\approx 1160 MeV, \ \ \ \ \ \ \    {\bf  \L(1380)}, 
  $$
  $$
   n=1,   \ \ \ E_{1}  \approx 1468 MeV, \ \ \ \ \ \ \ \ \    {\bf \L(1405)}, 
  $$
  $$
    n=2,   \ \ \ E_{2} \approx 1738 MeV,  \ \ \ \ \ \ \ \ \ \    {\bf \L(1670)}, 
  $$
     $$
   n=3,    \ \ \ E_{3} \approx 1983 MeV,  \ \ \ \ \  \ \ \ \ \    {\bf \L(1800)}, 
  $$
       $$
  n=4,    \ \ \ E_{3} \approx 2211 MeV,   \ \ \ \  \ \ \ \ \ \ \ \     {\bf \L(2000)}, 
  $$
      $$
  n=5,    \ \ \ E_{4}\approx 2426 MeV, (prediction) . \ \ \ \ \  
  $$
  \bigskip 
   \be
   model: \  \ell=2, \  \kappa=2; \  \L^0=(s|ud), \ J^P=3/2^+
   \ee
   $$
  CQD : K = 0.006, \ b = \frac{0.086}{\sqrt{MeV}}, \ \ \ \  
  $$
  $$
CQD(MeV ), \ \ \ \ \ \ \ \ \ \ \ \ \ \ \ \ \ \   {\bf data}, 
$$
   $$
   n=0, \ \  E_{0} \approx 1861 MeV, \ \ \ \ \ \ \ \ \ \ \ \ \    {\bf \L(1890)}.
  $$
  $$
    n=1,   \ \ \ E_{1} \approx 2128 MeV, \ \ \ \ \ \ \ \ \ \ \ \ \   {\bf \L(2070)}, 
  $$
 $$
   n=2,   \ \ \ E_{2} \approx 2371 MeV, (prediction). \ \ \ \ \ \ \ \ \ \ \ 
  $$
   \bigskip 
  \be
   model: \   \ell=1, \  \kappa=-2; \  \L^0=(s|ud), \ J^P=3/2^-
   \ee
   $$
  CQD : K = 0.006, \ b = \frac{0.068}{\sqrt{MeV}}, \ \ \ \ 
  $$
  $$
CQD(MeV ), \ \ \ \ \ \ \ \ \ \ \ \ \ \  {\bf data}, 
$$
   $$
     n=0, \ \  E_{0}\approx 1498 MeV, \ \ \ \ \ \ \ \ \   {\bf  \L(1520)}, 
  $$
  $$
   n=1,   \ \ \ E_{1}  \approx 1790 MeV, \ \ \ \ \ \ \ \ \ \     {\bf \L(1690)}, 
  $$
  $$
     n=2,   \ \ \ E_{2} \approx 2050 MeV, \ \ \ \ \  \ \ \ \ \ \    {\bf  \L(2050)}, 
  $$
   $$
    n=3,   \ \ \ E_{3} \approx 2289 MeV, \ \ \ \ \  \ \ \ \ \ \     {\bf  \L(2325)}, 
  $$
$$
    n=4,   \ \ \ E_{4} \approx 2512MeV, (prediction). \ \ \ \  
  $$
  \bigskip
  \be
   model:  \  \ell=2, \  \kappa=-3; \  \L^0=(s|ud), \ J^P=5/2^+
   \ee
   $$
  CQD : K = 0.006, \ b = \frac{0.082}{\sqrt{MeV}}, \ \ \ \ 
  $$
  $$
CQD(MeV ), \ \ \ \ \ \ \ \ \ \ \ \ \ \ \ \ \  {\bf data}, \ \ \ \ \ \ 
$$
   $$
    n=0, \ \  E_{0} \approx 1819 MeV, \ \ \ \ \ \ \ \     {\bf  \L(1820)},
  $$
  $$
    n=1,   \ \ \ E_{1}   \approx 2087MeV, \ \ \ \ \ \ \      {\bf  \L(2110)}, 
  $$
$$  
    n=2,   \ \ \ E_{2}   \approx  \ \ \   [2331`], \ \ \  
  $$
  \bigskip
  \be
   model: \ \ell=3, \  \kappa= 3; \    \L^0=(s|ud), \ J^P=5/2^-
   \ee
   $$
  CQD : K = 0.006, \ b = \frac{0.076}{\sqrt{MeV}}, \ \ \ \ 
  $$
  $$
CQD(MeV ), \ \ \ \ \ \ \ \ \ \ \ \ \ \ \ \ \ \ \     {\bf data}, 
$$
   $$
    n=0, \ \  E_{0} \approx 1874 MeV, \ \ \ \ \ \ \ \ \ \ \ \ \ \ \   {\bf \L(1830)}, 
  $$
  $$
    n=1,   \ \ \ E_{1} \approx 2134 MeV, \ \ \ \ \  \ \ \ \ \ \ \ \ \    {\bf \L(2080)}, 
  $$
$$
    n=2,   \ \ \ E_{2} \approx 2372 MeV, (prediction). \ \ \ \ \ \ \   
  $$
  \bigskip
  \be
   model: \  \ell=4,\   \kappa= 4; \    \L^0=(s|ud), \ J^P=7/2^+
   \ee
  $$
  CQD : K = 0.006, \ b = \frac{0.082}{\sqrt{MeV}}, \ \ 
  $$
  $$
CQD(MeV ), \ \ \ \ \ \ \ \ \ \ \ \ \ \ \ \ \     {\bf data}, 
$$
   $$
    n=0, \ \  E_{0} \approx 2075 MeV, \ \ \ \ \ \     {\bf  \L(2085)},
  $$
  $$
  n=1,   \ \ \ E_{1}   \approx 2321MeV, (prediction) .\ \ \ \  
  $$
  \bigskip
  \be
   model: \  \ell=3, \  \kappa= -4; \    \L^0=(s|ud), \ J^P=7/2^-
   \ee
   $$
  CQD : K = 0.006, \ b = \frac{0.092}{\sqrt{MeV}}, \ \ \ \  
  $$
  $$
CQD(MeV ), \ \ \ \ \ \ \ \ \ \ \ \ \ \ \ \ \ \ \  {\bf data}, 
$$
   $$
    n=0, \ \  E_{0}\approx 2077 MeV, \ \ \ \ \ \ \ \ \ \     {\bf  \L(2100)}, 
  $$
  $$
  n=1,   \ \  E_{1}  \approx 2328 MeV, (prediction) \ \ \ \ \ \ \ \ \ \ 
  $$
  \bigskip
  \be
   model: \ \ell=4, \  \kappa= -5; \    \L^0=(s|ud), \ J^P=9/2^+
   \ee
   $$
  CQD : K = 0.006, \ b = \frac{0.1}{\sqrt{MeV}}, \ \ \ \ 
  $$
  $$
CQD(MeV ), \ \ \ \ \ \ \ \ \ \ \ \ \ \ \  {\bf data}, 
$$
   $$
    n=0, \ \  E_{0} \approx 2302 MeV, \ \ \ \ \ \  {\bf  \L(2350)}, 
  $$
  $$
    n=1,   \ \ E_{1}  \approx 2541 MeV, (prediction) \ \ \ \ \   
  $$
  
  \subsection{$\S$ baryons}
  For $\S$ baryons $\S^+(uus), \ \S^0(uds), \ \S^-(dds)$ with positive charge and even parity, i.e., $N1/2^+$,the confining QD model gives the following results for $E_n\approx E_s(n) + E_{2u}(n)$ with the corresponding experimental values 
 ????
 \bigskip 
   \be
  model:  \  \ell=0, \  \kappa= -1; \     \S^+=(s|uu),  \ J^P=1/2^+
  \ee
  $$
  CQD : K = 0.006, \ b = \frac{0.055}{\sqrt{MeV}}, \ \ \ \ 
  $$
  $$
CQD(MeV ), \ \ \ \ \ \ \ \ \ \ \ \ \ \ \ \ \   {\bf data}, 
$$
  $$
   n=0, \ \  E_{0} \approx 1170 MeV, \ \ \ \ \ \ \ \ \    {\bf  \S(1189)}, 
  $$
  $$
  n=1,   \ \ E_{1}  \approx 1494 MeV, \ \ \ \ \ \ \ \ \ \ \   {\bf \S(1660)}, 
   $$
  $$
    n=2,   \ \ \ E_{2} \approx 1774 MeV, \ \ \ \ \  \ \ \ \ \    {\bf \S(1880)}, 
  $$
     $$
    n=3,    \ \ \ E_{3}\approx 2027 MeV, (prediction). \ \ \ \ \  \ \ \ \ \ 
  $$
\bigskip
  \be
   model: \ \ell=1, \  \kappa= 1; \     \S^+=(s|uu),  \ J^P=1/2^-
   \ee
   $$
  CQD : K = 0.006, \ b = \frac{0.07}{\sqrt{MeV}}, \ \ \ \  
  $$
  $$
CQD(MeV ), \ \ \ \ \ \ \ \ \ \ \ \ \ \ \ \ \ \  {\bf data}, 
$$
   $$
   n=0, \ \  E_{0} \approx 1518 MeV, \ \ \ \ \ \     {\bf \S(1620)},
  $$
  $$
   n=1,   \ \ \ E_{1}   \approx 1808 MeV, \ \ \ \ \ \ \   {\bf \S(1750)}, 
  $$
   $$
   n=2,   \ \ \ E_{2}  \approx 2069 MeV, \ \ \ \ \ \ \    {\bf \S(1900)}, 
  $$
  $$
    n=3,   \ \ \ E_{3} \approx 2307 MeV,  \ \ \ \ \ \ \ \ \   {\bf \S(2160)}, 
  $$
   $$
    n=4,   \ \ \ E_{4} \approx 2531 MeV, (prediction). \ \ \ \ \  
  $$
  \bigskip
  \be
   model: \ \ell=2, \  \kappa= 2; \    \S^+=(s|uu),  \ J^P=3/2^+    
   \ee
    $$
  CQD : K = 0.006, \ b = \frac{0.05}{\sqrt{MeV}}, \ \ \ \ 
  $$
  $$
CQD(MeV ), \ \ \ \ \ \ \ \ \ \ \ \ \ \ \ \ \ \   {\bf data}, 
$$
   $$
    n=0, \ \  E_{0} \approx 1370 MeV, \ \ \ \ \ \ \ \  \  {\bf  \S(1385)}, 
  $$
  $$
  n=1,   \ \ \ E_{1}  \approx 1659 MeV, \ \ \ \ \ \ \ \    {\bf  \S(1780)},
  $$
   $$
  n=2,   \ \ \ E_{2} \approx 1917MeV, \ \ \ \ \ \ \ \   {\bf \S(1940)}, 
  $$
   $$
  n=3,   \ \ \ E_{3}  \approx 2152 MeV, \ \ \ \ \ \ \ \ \ \  {\bf \S(2080)}, 
  $$
   $$
  n=4,   \ \ \ E_{4} \approx 2376 MeV, \ \ \ \ \ \ \ \    {\bf \S(2230)}, 
  $$
 $$
    n=5,   \ \ \ E_{5} \approx 2585 MeV, (prediction). okok\ \ \ \ \ \ 
  $$

\bigskip
  \be
   model: \ \ell=1, \  \kappa= -2; \    \S^+=(s|uu),  \ J^P= 3/2^-
   \ee
    $$
  CQD : K = 0.006,  \ b = \frac{0.065}{\sqrt{MeV}}, \ \ 
  $$
  $$
CQD(MeV ), \ \ \ \ \ \ \ \ \ \ \ \ \ \ \ \ \ \ \  {\bf data}, 
$$
   $$
    n=0,   \ \ \ E_{0}  \approx 1462 MeV, \ \ \ \ \  \ \ \   {\bf \S(1580)}, 
  $$
  $$
    n=1,   \ \ \ E_{2} \approx 1755 MeV, \ \ \ \ \  \ \ \     {\bf \S(1670)}, 
  $$
   $$
   n=2,   \ \ \ E_{3}\approx 2016 MeV, \ \ \ \ \  \ \ \   {\bf \S(1910)},
  $$
   $$
     n=3,   \ \ \ E_{4}\approx 2256 MeV, \ \ \ \ \  \ \ \    {\bf \S(2010)}, 
  $$
$$
   n=4,   \ \ \ E_{5}(3/2^-) \approx 2480 MeV, (prediction). \ \ \ \ \ 
  $$
   \bigskip 
  \be
   model:\  \ell=2,\   \kappa= -3; \       \S^+=(s|uu),  \ J^P=5/2^+
  \ee
   $$
  CQD : K = 0.006, \ b = \frac{0.085}{\sqrt{MeV}}, \ \ \ \  
  $$
  $$
CQD(MeV ), \ \ \ \ \ \ \ \ \ \ \ \ \ \ \ \ \   {\bf data}, 
$$
  $$
   n=0, \ \  E_{0}\approx 1856 MeV, \ \ \ \ \ \ \ \ \ \  {\bf  \S(1915)}, 
  $$
  $$
    n=1,   \ \ \ E_{1}  \approx 2122 MeV, \ \ \ \ \ \ \ \ \   {\bf \S(2070)}, 
  $$
$$
   n=2,   \ \ \ E_{2}  \approx 2366 MeV, (prediction). \ \ \ \ \ \ \ 
  $$
 \bigskip
  \be
   model:\  \ell=3, \  \kappa= 3; \    \S^+=(s|uu),  \ J^P=5/2^-
   \ee
    $$
  CQD : K = 0.006, \ b = \frac{0.065}{\sqrt{MeV}}, 
  $$
  $$
CQD(MeV ), \ \ \ \ \ \ \ \ \ \ \ \ \ \  {\bf data}, 
$$
   $$
    n=0, \ \  E_{0} \approx 1746 MeV, \ \ \ \     {\bf  \S(1775)}, 
  $$
$$
   n=1,   \ \ \ E_{1}  \approx 2009 MeV, (prediction). \ \ \ \ 
  $$
 \bigskip
  \be
   model: \ \ell=4,  \ \kappa= 4; \     \S^+=(s|uu),  \ J^P= 7/2^+  
   \ee
    $$
  CQD : K = 0.006,  \ b = \frac{0.09}{\sqrt{MeV}}, 
  $$
  $$
CQD(MeV ), \ \ \ \ \ \ \ \ \ \ \ \     {\bf data}, 
$$
   $$
    n=0, \ \  E_{0} \approx 2053 MeV, \ \ \    {\bf  \S(2030)}, 
  $$
  $$
   n=1,   \ \  E_{1}  \approx 2305 MeV, (prediction). \ \ \ \ \ 
 $$
 
   \subsection{${\bf Omega}$ and charmed  baryons}
  For $\Om^-(s|ss)$ and charmed baryons $\L_c^+(c|ud)$, the confining QD model gives the following results for $E_n\approx E_s(n) + E_{2s}(n)$ with the corresponding experimental values.\cite{3} 
\bigskip

\noindent
{\bf Omega  baryons}  
\bigskip
   \be
  model:  \  \ell=0, \  \kappa= -1; \ \ \ \ \ \   \Om^-=(s|ss), \ \  J^P=1/2^+
  \ee
  $$
  CQD : K = 0.006, \ \ b = \frac{0.08}{\sqrt{MeV}}, \ \ \ \ 
  $$
  $$
CQD(MeV ), \ \ \ \ \ \ \ \ \ \ \ \ \ \ \ \ \ \ \   {\bf data},  
$$
  $$
   n=0, \ \  E_{0} \approx 1660 MeV, \ \ \ \ \ \ \ \ \   {\bf  \Om(1672)}, 
  $$
  $$
  n=1,   \ \ E_{1}  \approx 1961 MeV, \ \ \ \ \ \ \ \ \ \   {\bf \Om(2012)}, 
   $$
  $$
    n=2,   \ \ \ E_{2} \approx 2227 MeV,   \ \ \ \ \ \ \ \   {\bf \Om(2250)}, 
  $$
  $$
  n=3,   \ \ E_{3}  \approx 2470 MeV, \ \ \ \ \ \ \ \ \ \   {\bf \Om(2470)}, 
   $$
  $$
    n=4,   \ \ \ E_{2} \approx 2696 MeV, (prediction).  \ \ \ \ \ 
  $$
  \bigskip  
  \be
  model:  \  \ell=1, \  \kappa= -2; \   \Om^-=(s|ss), \   J^P=3/2^- 
  \ee
  $$
  CQD : K = 0.006, \ \ b = \frac{0.11}{\sqrt{MeV}}, \ \ \ \ 
  $$
  $$
CQD(MeV ), \ \ \ \ \ \ \ \ \ \ \ \ \ \ \ \ \ \ \  {\bf data},  
$$
  $$
   n=0, \ \  E_{0} \approx 2195 MeV, \ \ \ \ \ \ \ \ \    {\bf  \Om(2250)}, 
  $$
  $$
    n=1,   \ \ \ E_{1} \approx 2460 MeV, (prediction). \ \ \    
  $$

\bigskip
\noindent
{\bf Charmed baryons}
\bigskip
\be
  model:  \  \ell=0, \  \kappa= -1; \ \ \ \ \ \   \L^+_c=(c|ud), \ \  J^P=1/2^+
  \ee
  $$
  CQD : K = 0.006, \ \ b = \frac{0.06}{\sqrt{MeV}}, \ \ \ \ 
  $$
  $$
CQD(MeV ), \ \ \ \ \ \ \ \ \ \ \ \ \ \ \ \ \   {\bf data},  
$$
  $$
   n=0, \ \  E_{0} \approx 2299 MeV, \ \ \ \ \ \ \    {\bf  \L_c(2286)}, 
  $$
  $$
    n=1,   \ \ \ E_{1} \approx 2524 MeV, (prediction). \ \ \ \ \  \ \ \    
  $$
\bigskip
\be
  model:  \  \ell=1, \  \kappa= 1; \ \ \ \ \ \   \L_c=(c|ud), \ \  J^P=1/2^- 
  \ee
  $$
  CQD : K = 0.006, \ \ b = \frac{0.075}{\sqrt{MeV}}, \ \ \ \ 
  $$
  $$
CQD(MeV ), \ \ \ \ \ \ \ \ \ \ \ \ \ \ \ \ \ \  {\bf data},  
$$
  $$
   n=0, \ \  E_{0} \approx 2611 MeV, \ \ \ \ \ \ \ \ \   {\bf  \L_c(2595)}, 
  $$
  $$
    n=1,   \ \ \ E_{1} \approx 2825 MeV, (prediction). \ \ \ \ \  \ \ \  
  $$
\be
  model:  \  \ell=1, \  \kappa= -2; \ \ \ \ \ \   \L_c=(c|ud), \ \  J^P=3/2^- 
  \ee
  $$
  CQD : K = 0.006, \ \ b = \frac{0.08}{\sqrt{MeV}}, \ \ \ \ 
  $$
  $$
CQD(MeV ), \ \ \ \ \ \ \ \ \ \ \ \ \ \ \ \ \ \ \  {\bf data},  
$$
  $$
   n=0, \ \  E_{0} \approx 2684 MeV, \ \ \ \ \ \ \ \   {\bf  \L_c(2625)}, 
  $$
   $$
   n=1, \ \  E_{1} \approx 2896 MeV, \ \ \ \ \ \ \ \  {\bf  \L_c(2940)}, 
  $$
  $$
    n=2,   \ \ \ E_{2} \approx 3099 MeV, (prediction). \ \ \ \ \  
  $$ 
 \bigskip
\be
  model:  \  \ell=2, \  \kappa= 2; \ \ \ \ \ \   \L_c=(c|ud), \ \  J^P=3/2^+
  \ee
  $$
  CQD : K = 0.006, \ \ b = \frac{0.085}{\sqrt{MeV}}, \ \ \ \ 
  $$
  $$
CQD(MeV ), \ \ \ \ \ \ \ \ \ \ \ \ \ \ \ \ \ \  {\bf data},  
$$
  $$
   n=0, \ \  E_{0} \approx 2851 MeV, \ \ \ \ \ \ \ \    {\bf  \L_c(2860)}, 
  $$
  $$
    n=1,   \ \ \ E_{1} \approx 3058 MeV, (prediction). \ \ \ \ \    
  $$
 \bigskip
\be
  model:  \  \ell=0, \  \kappa= -1; \ \ \ \ \ \   \S_c=(c|ud), \ \  J^P=1/2^+ 
  \ee
  $$
  CQD : K = 0.006, \ \ b = \frac{0.07}{\sqrt{MeV}}, \ \ \ \ 
  $$
  $$
CQD(MeV ), \ \ \ \ \ \ \ \ \ \ \ \ \ \ \ \ \  {\bf data},  
$$
  $$
   n=0, \ \  E_{0} \approx 2433 MeV, \ \ \ \ \ \ \ \ \   {\bf  \S_c(2456)}, 
  $$
  $$
    n=1,   \ \ \ E_{1} \approx 2655 MeV, (prediction).\ \ \ \ \    
  $$  
\bigskip
\be
  model:  \  \ell=2, \  \kappa= 2; \ \ \ \ \ \   \S_c=(c|ud), \ \  J^P=3/2^+
  \ee
  $$
  CQD : K = 0.006, \ \ b = \frac{0.06}{\sqrt{MeV}}, \ \ \ \ 
  $$
  $$
CQD(MeV ), \ \ \ \ \ \ \ \ \ \ \ \ \ \ \ \ \ \   {\bf data},  
$$
  $$
   n=0, \ \  E_{0} \approx 2519 MeV, \ \ \ \ \ \ \ \ \  {\bf  \S_c(2520)}, 
  $$
  $$
    n=1,   \ \ \ E_{1} \approx 2733 MeV, (prediction). \ \ \ \ \  \ \ \    
  $$ 
 \bigskip
\be
  model:  \  \ell=2, \  \kappa= -3; \ \ \ \ \ \   \S_c=(c|ud), \ \  J^P=5/2^+
  \ee
  $$
  CQD : K = 0.006, \ \ b = \frac{0.08}{\sqrt{MeV}}, \ \ \ \ 
  $$
  $$
CQD(MeV ), \ \ \ \ \ \ \ \ \ \ \ \ \ \ \ \ \ \ \  {\bf data},  
$$
  $$
   n=0, \ \  E_{0} \approx 2793 MeV, \ \ \ \ \ \ \ \ \   {\bf  \S_c(2800)}, 
  $$
  $$
    n=1,   \ \ \ E_{1} \approx 3000 MeV, (prediction). \ \ \ \ \  \ \   
  $$ 
  \bigskip
   
  \bigskip
  
\noindent
{\bf $\Xi$ baryons}
\bigskip
\be
  model:  \  \ell=0, \  \kappa= -1; \ \ \ \ \ \   \Xi^+_c=(c|us), \ \  J^P=1/2^+
  \ee
  $$
  CQD : K = 0.006, \ \ b = \frac{0.057}{\sqrt{MeV}}, \ \ \ \ 
  $$
  $$
CQD(MeV ), \ \ \ \ \ \ \ \ \ \ \ \ \ \ \ \ \   {\bf data},  
$$
  $$
   n=0, \ \  E_{0} \approx 2473 MeV, \ \ \ \ \ \ \    {\bf  \Xi_c(2468)}, 
  $$
  $$
    n=1,   \ \ \ E_{1} \approx 2721 MeV, \ \ \ \ \  \ \ \    {\bf \Xi_c(2578)}. 
  $$
   $$
    n=2,   \ \ \ E_{2} \approx 2959 MeV, (prediction) \ \ \ \ \  \ \ \   
  $$
  \bigskip
\be
  model:  \  \ell=1, \  \kappa= 1; \ \ \ \ \ \   \Xi_c=(c|us), \ \  J^P=1/2^- 
  \ee
  $$
  CQD : K = 0.006, \ \ b = \frac{0.08}{\sqrt{MeV}}, \ \ \ \ 
  $$
  $$
CQD(MeV ), \ \ \ \ \ \ \ \ \ \ \ \ \ \ \ \ \ \  {\bf data},  
$$
  $$
   n=0, \ \  E_{0} \approx 2764 MeV, \ \ \ \ \ \ \ \ \   {\bf  \Xi_c(2790)}, 
  $$
  $$
    n=1,   \ \ \ E_{1} \approx 2890 MeV, (prediction). \ \ \ \ \  \ \ \  
  $$
\bigskip
\be
  model:  \  \ell=2, \  \kappa= 2; \ \ \ \ \ \   \Xi_c=(c|us), \ \  J^P=3/2^+
  \ee
  $$
  CQD : K = 0.006, \ \ b = \frac{0.067}{\sqrt{MeV}}, \ \ \ \ 
  $$
 $$
CQD(MeV ), \ \ \ \ \ \ \ \ \ \ \ \ \ \ \ \ \ \  {\bf data},  
$$
  $$
   n=0, \ \  E_{0} \approx 2698 MeV, \ \ \ \ \ \ \ \ \   {\bf  \Xi_c(2645)}, 
  $$
  $$
    n=1,   \ \ \ E_{1} \approx 2905 MeV, (prediction). \ \ \ \ \  \ \ \  
  $$
\bigskip
\be
  model:  \  \ell=1, \  \kappa= -2; \ \ \ \ \ \   \Xi_c=(c|us), \ \  J^P=3/2^-
  \ee
  $$
  CQD : K = 0.006, \ \ b = \frac{0.083}{\sqrt{MeV}}, \ \ \ \ 
  $$  
  $$
   n=0, \ \  E_{0} \approx 2805 MeV, \ \ \ \ \ \ \ \ \   {\bf  \Xi_c(2815)}, 
  $$
  $$
    n=1,   \ \ \ E_{1} \approx 3015 MeV, (prediction). \ \ \ \ \  \ \ \  
  $$
  \bigskip
\be
  model:  \  \ell=0, \  \kappa= -1; \ \ \ \ \ \   \Om_c=(c|ss), \ \  J^P=1/2^+
  \ee
  $$
  CQD : K = 0.006, \ \ b = \frac{0.08}{\sqrt{MeV}}, \ \ \ \ 
  $$
 $$
CQD(MeV ), \ \ \ \ \ \ \ \ \ \ \ \ \ \ \ \ \ \  {\bf data},  
$$
  $$
   n=0, \ \  E_{0} \approx 2742 MeV, \ \ \ \ \ \ \ \ \   {\bf  \Om_c(2700)}, 
  $$
  $$
   n=1, \ \  E_{1} \approx 2957 MeV, \ \ \ \ \ \ \ \ \   {\bf  \Om_c(3000)}, 
  $$
  $$
    n=2,   \ \ \ E_{2} \approx 3162 MeV, (prediction). \ \ \ \ \  \ \ \  
  $$
\bigskip
\be
  model:  \  \ell=2, \  \kappa= 2; \ \ \ \ \ \   \Xi_c=(c|ss), \ \  J^P=3/2^+
  \ee
  $$
  CQD : K = 0.006, \ \ b = \frac{0.065}{\sqrt{MeV}}, \ \ \ \ 
  $$
  $$
CQD(MeV ), \ \ \ \ \ \ \ \ \ \ \ \ \ \ \ \ \ \  {\bf data},  
$$
  $$
   n=0, \ \  E_{0} \approx 2757 MeV, \ \ \ \ \ \ \ \ \   {\bf  \Om_c(2770)}, 
  $$
  $$
   n=1, \ \  E_{1} \approx 2967 MeV, \ \ \ \ \ \    {\bf  \Om_c(3000)}, (assumed \ J^P)
  $$
  $$
    n=2,   \ \ \ E_{2} \approx 3167 MeV, (prediction). \ \ \ \ \  \ \ \  
  $$
   \bigskip
\noindent
\subsection{{\bf Bottom baryons}}
\be
  model:  \  \ell=0, \  \kappa= -1; \ \ \ \ \ \   \Ld^+_b=(b|ud), \ \  J^P=1/2^+
  \ee
  $$
  CQD : K = 0.006, \ \ b = \frac{0.093}{\sqrt{MeV}}, \ \ \ \ 
  $$
  $$
CQD(MeV ), \ \ \ \ \ \ \ \ \ \ \ \ \ \ \ \ \   {\bf data},  
$$
  $$
   n=0, \ \  E_{0} \approx 5600 MeV, \ \ \ \ \ \ \    {\bf  \Ld_b^0(5620)}, 
  $$
  $$
    n=1,   \ \ \ E_{1} \approx 5742 MeV, (prediction) \ \ \ \ \  \ \ \  
  $$
\be
  model:  \  \ell=1, \  \kappa= 1; \ \ \ \ \ \   \Xi_b=(b|ud), \ \  J^P=1/2^- 
  \ee
  $$
  CQD : K = 0.006, \ \ b = \frac{0.11}{\sqrt{MeV}}, \ \ \ \ 
  $$
  $$
CQD(MeV ), \ \ \ \ \ \ \ \ \ \ \ \ \ \ \ \ \ \  {\bf data},  
$$
  $$
   n=0, \ \  E_{0} \approx 5903 MeV, \ \ \ \ \ \ \ \ \   {\bf  \Ld_b(5912)}, 
  $$
  $$
    n=1,   \ \ \ E_{1} \approx 6043 MeV, (prediction). \ \ \ \ \  \ \ \  
  $$
\be
  model:  \  \ell=1, \  \kappa= -2; \ \ \ \ \ \   \Xi_b=(b|ud), \ \  J^P=3/2^-
  \ee
  $$
  CQD : K = 0.006, \ \ b = \frac{0.11}{\sqrt{MeV}}, \ \ \ \ 
  $$
 $$
CQD(MeV ), \ \ \ \ \ \ \ \ \ \ \ \ \ \ \ \ \ \  {\bf data},  
$$
  $$
   n=0, \ \  E_{0} \approx 5908 MeV, \ \ \ \ \ \ \ \ \   {\bf  \Ld_b(5920)}, 
  $$
  $$
    n=1,   \ \ \ E_{1} \approx 6048 MeV, (prediction). \ \ \ \ \  \ \ \  
  $$
\be
  model:  \  \ell=2, \  \kappa= 2; \ \ \ \ \ \   \Xi_b=(b|ud), \ \  J^P=3/2^+
  \ee
  $$
  CQD : K = 0.006, \ \ b = \frac{0.12}{\sqrt{MeV}}, \ \ \ \ 
  $$  
  $$
   n=0, \ \  E_{0} \approx 6109 MeV, \ \ \ \ \ \ \ \ \   {\bf  \Ld_b(6146)}, 
  $$
  $$
    n=1,   \ \ \ E_{1} \approx 6248 MeV, (prediction). \ \ \ \ \  \ \ \  
  $$
\be
  model:  \  \ell=2, \  \kappa= -3; \ \ \ \ \ \   \Ld_b=(b|ud), \ \  J^P=5/2^+
  \ee
  $$
  CQD : K = 0.006, \ \ b = \frac{0.12}{\sqrt{MeV}}, \ \ \ \ 
  $$
 $$
CQD(MeV ), \ \ \ \ \ \ \ \ \ \ \ \ \ \ \ \ \ \  {\bf data},  
$$
  $$
   n=0, \ \  E_{0} \approx 6117 MeV, \ \ \ \ \ \ \ \ \   {\bf  \Ld_b(6152)}, 
  $$
  $$
    n=1,   \ \ \ E_{2} \approx 6255 MeV, (prediction). \ \ \ \ \  \ \ \  
  $$
\be
  model:  \  \ell=0, \  \kappa= -1; \ \ \ \ \ \   \Xi_b=(b|ds), \ \  J^P=1/2^+
  \ee
  $$
  CQD : K = 0.006, \ \ b = \frac{0.1}{\sqrt{MeV}}, \ \ \ \ 
  $$
  $$
CQD(MeV ), \ \ \ \ \ \ \ \ \ \ \ \ \ \ \ \ \ \  {\bf data},  
$$
  $$
   n=0, \ \  E_{0} \approx 5781 MeV, \ \ \ \ \ \ \ \ \   {\bf  \Xi^0_b(5800)}, 
  $$
  $$
   n=1, \ \  E_{1} \approx 5940 MeV, \ \ \ \ \ \ \ \ \   {\bf  \Xi'_b(5935)}, 
  $$
  $$
    n=2,   \ \ \ E_{2} \approx 6060 MeV, (prediction). \ \ \ \ \  \ \ \  
  $$
\be
  model:  \  \ell=2, \  \kappa= 2; \ \ \ \ \ \   \Xi_b=(b|ds), \ \  J^P=3/2^+
  \ee
  $$
  CQD : K = 0.006, \ \ b = \frac{0.1}{\sqrt{MeV}}, \ \ \ \ 
  $$
  $$
CQD(MeV ), \ \ \ \ \ \ \ \ \ \ \ \ \ \ \ \ \ \  {\bf data},  
$$
  $$
   n=0, \ \  E_{0} \approx 5919 MeV, \ \ \ \ \ \ \ \ \   {\bf  \Xi_b(5945)^-}, 
  $$
  $$
   n=1, \ \  E_{1} \approx 6056 MeV, (prediction) \ \ \ \ \ \ \ \ \  
  $$

So far, roughly 120 baryons with masses from $\approx$1000MeV to $\approx$6000 MeV are calculated in the CQD model and compared with data.  We have seen a reasonably ageement. 
 
\section{The neutron-proton mass difference and different nucleon core structures}


 The confining QD model provides an interesting tool to calculate and understand the neutron-proton mass difference.  According to this model, all 3-quark baryons with neutral quark charges can be pictured as having a `core' quark at the center and two quarks in the surrounding quantum cloud shell.   
For comparisons, let us consider three  baryon mass differences:
\bigskip

(I) Mass difference of neutron-proton:  

Since the proton $p(d|uu)$ has two u quarks in the quantum shell with the s state (total spin 0) and a d quark at the core.  The neutron $n(d|ud)$ has u and d quarks in the shell and one d quark at the core.  In this model, the d quark mass is 4.67 MeV.  However, once the d quark is at the proton core, its effective mass is about 900MeV (due to its interactions with via quark Hooke potential $V_{qH}$).  In contrast, the d quark in the quantum shell only has approximately 10 MeV (due to its interaction via the confining potential C(r)), as shown by $E_d$ and $E_{du}/2$ in (91) below. 

Based on (24)-(26) with K=0.012 and b=0.0212, the proton and the neutron masses are given by
\be
m(p^+) = E_d +E_{uu}\approx  938.62 MeV, \ \  
\ee
$$
m(n^0)= E_d + E_{du}\approx 939.19 MeV,
$$
$$
E_d= 911.85 MeV, \ \  E_u=E_{uu}/2\approx E_{du}/2 \approx 13.4 MeV,
$$
    The neutron-proton mass difference is given by
\be
m(n^0) - m(p^+) \approx  0.57 MeV  > 0.
\ee
\be
[m(n^0) - m(p^+) ]_{data} \approx  1.3 MeV  > 0.
\ee
In view of the approximate nature of the model, the neutron-proton mass difference (92) appears to be in reasonable agreement with  the experimental value $\approx +1.3 MeV.$   
\bigskip

(II) Mass difference of $\S^+, \S^0$ and $\S^- $ in an isotriplet: 

Let us consider the mass of $\S^+(s|uu).$  Similar to (62), we use K=0.006 and b=0.057/$\sqrt{MeV}$, we have m($\S^+$)= 1195.4 MeV.  Similarly, we calculate the energy eigenvalues of $\S^0(s|ud)$ and $\S^-(s|dd)$.  The rresults of the confining QD model are listed below and compared with data:
\be
  CQD : K = 0.006, \ \ b = \frac{0.057}{\sqrt{MeV}},  \ \  \ \  J^P=1/2^+ \ \ \ \ 
  \ee
  $$
  \ell=0, \  \kappa= -1; \ \ \ \ \ \ \ \ \  {\bf data},  
$$
  $$
   n=0, \ \ \S^+(s|uu)\approx 1195.4 MeV, \ \ \ \ \ \ \ \ \   {\bf  \S^+(1189.4)}, 
  $$
  $$
   n=0, \ \   \S^+(s|ud) \approx  1196.5 MeV, \ \ \ \ \ \ \ \ \   {\bf  \S^+(1192.6)},   
  $$
 $$
   n=0, \ \    \S^+(s|dd)\approx  1197.7 MeV \ \ \ \ \ \ \ \ \   {\bf  \S^+(1197.5)},  
  $$
\bigskip

(III)  Mass difference of charmed $\Xi_c$ baryons in two isodoublets:  

We consider the mass difference of [$\Xi_c^+(c|su),  \ \Xi_c^0(c|ds)$] and of  [$\Xi_c^{'+}(c|su),  \ \Xi_c^{'0}(c|ds)$].   Similar to (78), we use K=0.006 and b=0.06/$\sqrt{MeV}$, we have m($\Xi_c^+$)= 2472.7 MeV.  Similarly, we calculate the energy eigenvalues of other charmed baryons $\Xi_c$.  The results of the confining QD model for mass differences of charmed baryons in two isodoublets are listed below and compared with data:
\be
  CQD : K = 0.006, \ \ b = \frac{0.057}{\sqrt{MeV}},  \ \ \ \ \ \    \ \  J^P=1/2^+ \ \ \ \ 
  \ee
  $$
  \ell=0, \  \kappa= -1; \ \ \ \ \ \ \ \ \ \ \ \ \  {\bf data},  
$$
  $$
   n=0, \ \ \Xi_c^+(c|su)\approx 2472.7   MeV, \ \ \ \     {\bf  \Xi_c^+(2467.9)}, 
  $$
  $$
   n=0, \ \   \Xi_c^0(c|sd) \approx 2473.7  MeV, \ \ \ \ \    {\bf  \Xi_c^0(2470.9)},   
   $$
   $$
   m(\Xi_c^0) - m(\Xi_c^+) \approx  1.0 MeV, \ \ \ \  
   $$
   $$
   [  m(\Xi_c^0) - m(\Xi_c^+)]_{data} \approx 3.0 MeV.
   $$
   
   \be
  CQD : K = 0.006, \ \ b = \frac{0.057}{\sqrt{MeV}},  \ \ \ \ \ \    \ \  J^P=1/2^+ \ \ \ \ 
  \ee
 $$
   n=1, \ \    \Xi_c^{'+}(c|su)\approx 2723.7   MeV \ \ \ \ \ \ \ \ \   {\bf  \Xi_c^{'+}(2578.4)},  
  $$
  $$
  n=1, \ \   \Xi_c^{'0}(c|sd) \approx 2724.1  MeV, \ \ \ \ \ \ \ \ \   {\bf  \Xi_c^{'0}(2579.2)},  
  $$
  $$
   m(\Xi_c^{'0}) - m(\Xi_c^{'+}) \approx  0.4 MeV, \ \ \ \  
   $$
   $$
   [  m(\Xi_c^{'0}) - m(\Xi_c^{'+})]_{data} \approx 0.8 MeV.
   $$
  
In view of the approximate nature of the confining QD model, the sub-models (I)-(III) give reasonable agreements with the experimental values of  the baryon mass differences, as shown in  (92)-(96).

 
\section{`Proton spin crisis' and experimental tests for the new proton core structure}
 The Ashman $et \ al$ experiment of deep inelastic muon-proton experiment\cite{12} revealed that quarks contribute surprisingly little to the proton's spin. ``This ‘spin crisis’ is doing much to clarify the subtle departures of the underlying field theory from the naive quark model."\cite{13}  In the naive quark model, color charges are assumed to allow multiple quarks to be in the same quantum state, say, in a baryon without violating the exculsion principle.
 
  However, the present confining QD model for a baryon with a core quark and two quarks in the surrounding quantum shell, there is no `preconceived difficulty' of multiple quarks to be in the same quantum state.   Furthermore, the confining QD model can give a reasonable understanding of roughly 120 baryon masses with two potentials, as shown in (28)-(96), and without having to assume the 3 color charges.  We note that the agreement between the model and the baryon mass spectra appears to be insensitive to whether one assumes 3 color charge for quarks or not.
 
In the muon-proton scattering, only the electric charges of muons and protons are directly involved, while color charges are not directly involved.  In this aspect, the proton core d quark (with -e/3) and two surrounding u quarks (with 2[2e/3]) behave similarly to a helium atoms with two electrons in the s state.  Thus, the sub-model (I) in sec. 6 with $p^+(d|uu)$ and (78)-(79) appears to be consistent and supported by the deep inelastic muon-proton scatterings, the baryon mass spectra and the neutron-proton mass difference, as demonstrated in section 6.   
 
  The confining QD model implies that the spin properties of the proton are not carried by the two u quarks in the surrouding quantum shell.  Rather, the proton spin are carried by the d quark hidden at the proton core.  The confining QD model predicts that when the energy of the incoming muon increases, the deep inelastic muon-proton scattering results will show more and more contribution to the proton spin by the d quark with the electric charge -e/3 at the `core' of the proton.

 We propose that deep inelastic muon-proton or electron-proton scatterings with higher energies can be used to test further the confining QD model, which predicts that the proton has core structure with the d quark with spin 1/2 and the electric cxharge -e/3.  Such experiments for testing the proton structure could play a role similar to that of the Rutherfoord experiments, which reveals the atomic structure. 
  
  \section{Cosmological implications of the confining QD model with new mass source}
 
 The confining QD model explores basic quarks and antiquarks interactions and the formations of  baryon and anti-baryon resonance states, which are closely related to matter and antimatter in the Big-Jets model for the birth of the universe.\cite{14} 

So far, we do not have ideas for understanding the origin of masses of fundamental quarks and leptons.  However, the confining QD model reveals a new mechanism of generating mass by quark's couplings with the confining potential C(r) and the quark Hooke potential $V_{qH}$.  This new mechanism for generating mass has interesting and significant cosmological implications.   

Suppose the beginning of the universe was the creation of two Big-Jets\cite{14} rather than a big bang.  The Big-Jets model mplies that 
 the universe begins with two big jets moving in oppose directions.  Each jet may be considered as a gigantic fire ball (similar to a big bang), which contains enormous numbers of quarks and anti-quarks, etc. with high energies.   After complicated collisions and decays, if one fire ball turned out to be dominated by matter such as protons, neutrons, electrons; then the other fireball must be dominated by corresponding antimatter such and anti-protons, anti-neutrons etc., based on the CPT invariance in particle physics.  The confining QD model implies a new mass source for all these particles and anti-particles made of quarks and anti-quarks.\footnote{ After billions of years, the matter half-universe cools down to a blackbody with 3K, and the antimatter half-universe must also cools down to an anti-blackbody with 3 K.\cite{14} The Big Jets model explains why we do not observe anti-stars and anti-galaxies in our observable matter half-universe. }  

For cosmological implications of the new mass source, suppose the present total observable mass of our matter half-universe is $M_{hu}$.\footnote{Here, we do not consider dark matter and dark energy, which were discussed elsewhere. See chs. 2-3 in ref. 5}  The confining QD model implies that at the very begining, the total mass of quarks is about $0.01M_{hu}. $ Through the strong interactions, quarks formed all kinds of resonance states of baryons and mesons.  The baryon resonances with large masses would decay to resonances with smaller masses.  The final decay products are protons, neutrons, etc.  The decay lifetimes of resonances are in the range of  $10^{-10} - 10^{-25}$ sec.  Thus, within one second after the Big-Jets event, almost all mass of the matter (or antimatter) half-universe would be carried by protons and neutrons (or anti-protons and anti-neutrons).  In such a short time, the observable mass of our matter half-universe increases from about $0.01M_{hu}$ to $1 M_{hu}$.   Let us estimate the increase of mass within the confining model:  Based on the result (28), the model gives the proton mass
\be
E_0=E_d(0)+ E_{2u}(0)\approx 920MeV,
\ee
$$
E_d(0)\approx 893\approx 888.3 + m_d, \ \ \ \  E_{2u}(0)\approx 27 \approx 22.7 +2m_u,
$$
where $m_d=4.67 MeV$ and $m_u= 2.16 MeV.$. This estimation hold also for the neutron.\footnote{The neutron mass is given by $E_0=E_d(0) + E_{ud}(0)$, where $E_{ud}(0)\approx E_{2u}(0)$ and one has the same estimation.}   Thus, the new mass source contributes roughly  $ (888.3+22.7)/920 \approx 99\%$ through protons and neutrons to the observable mass in the physical universe.   

It is interesting to compare this mass source based on confining quarkdynamics with that of Mach's principle in classical physics.  Mach's principle proposed that the inertial properties of the body are due to its interactions with all other objects in the universe.  But so far, there is neither demonstration of specific interactions based on a model nor experimental data to support Mach's principle.  In contrast,
  the confining QD model implys that $\approx  99\% $ of the observable mass of the physical universe is originated from the core d quark's strong interactions via quark Hooke potential $V_{qH}$ and from the two shell u quarks' interactions via confining potential C(r).  All these interactions take place completely inside the protons and the neutrons and are associated with specific potentials $V_{qH}$ and C(r).   The new mass source is suppored by the baryon mass spectra data, as shown in (28)-(90).
  
   The Big Jets model based on the CPT theorem implies that there exists an antimatter half-universe, which was created and evolved similarly to our matter half-universe.  The CPT theorem and the coninfing quark model imply that about 99 \% of the masses in the matter half-universes are originated from the strong 3-quark interactions inside protons and neutrons.  Also, about 99\% of the masses in the antimatter half-universe are originated from the strong 3-antiquark interacitons inside antiproton and antineutrons.

\section{Discussion and conclusion}
The coupling constant Q in the quark Hooke's potential is not independent of the coupling constant $Q'$ in the confining potential, as shown in (25).  The reason is that the quark Hooke's potential is produced by the two quarks with the confining potentials in the quantum shell surrounding the baryon core.  The model assumes that $V_o$ (or b) has the same value for a specific sub-spectrum specified by $J^P$.  A sub-spectrum may involve 1 to 6 eigenstates of baryons, as one can see in (28)-(96).   

Suppose there is only one kind of quark charge $q_o$, the confining model can also give a reasonable understanding of roughly 120 baryon masses based on two potentials.  In this case, the ratio of quark charge $q_o$ to the color charge $q_c$ is $\sqrt{3}$.    In other words, the agreement between the confining model and the baryon mass spectra holds  for both cases: (A) there are 3 color charges, $q_c$, for quarks and (B) there is only one kind of quark charge $q_o$, where $q_o=\sqrt{3} \ q_c.$ 

The general Yang-Mills symmetry also dictates the invariant Lagranian to have the renormalizable form\cite{1}
\be
L_{gYM}=\frac{L^2_s}{2}[\p^\mu H^a_{\mu\ld} \p_{\nu} H^{\a\nu\ld}+\p^\mu B_{\mu\ld} \p_{\nu} B^{\nu \ld}]
\ee
$$
\ov{q}[ig^\mu(\p_\mu -ig_s H^\a_\mu \frac{\ld^\a}{2} -ig_b B_\mu) -M]q.
$$
As a rsult, the usual results in QD will not be upset by the higher order corrections based on (98) with general Yang-Mills symmetry.\cite{15,16,17}

It is interesting and intriguing that there are 6 quarks and 6 leptons at the present time. If the correspondence between 6 quarks and 6 leptons is not purely by chance, then it is possible that there is a deeper reason behind it.\cite{16} We shall consider this possiblity, i..e., an idealized quark-lepton mass symmetry, in the Appendix based on the confining QD model.
  
The small neutron-proton mass difference of roughly 1 MeV has been a long standing problem in particle physics.  It was investigated in the 1960s based on a modified  pion-nucleon dispersion relations.  Based on a modified pion-nucleon dispersion relations, the consistency between the numerical results and the experimental measurement is greatly improved.  However, the results are very sensitive to the precise choice of parameters in the modified dispersion relations.\cite{18,19}   It is gratefying that the long standing problem of neutron-proton mass difference could also be reasonably understood by the confining QD model with a specific d quark as the cores of proton and neutron, as discussed in sec. 6.  Since we did not aim at the best fit for the baryon spectra, hence, it is possible to adjust K and b to get a better fit of the baryon data in (28)-(96).
 
 In conclusion, the confining QD model postulates that any baryon has 
 a core quark and a surrounding quantum shell with 2 quarks. 
 Each quark obeys the Sonine-Laguerre equation based on the relaivistic Hamiltonian with a linear  potential or the quark Hooke potential.  Essentially,  two coupling constants, i.e., K=0.012  and K=0.006, and one parameter b=$V_o/4\sqrt{Q} $ for each sub-spectrum suffice to understand roughly 120 baryon masses, approximately within 20\% in percent deviation.   The model reveals a new mass source for baryons, including protons and neutrons, which are omnipresent in the observable universe.  Consequently, about 99\% of the observable mass in the universe is originated through the d quark interaction with a large coupling constant in the quark Hooke potential $V_{qH}$ and confining linear potential C(r).  Based on the CPT theorem, this new mechanism for generating mass must hold also in the antimatter half-universe  with omnipresent antiprotons and antineutrons. 
 
The work was supported in part by the Jing Shin Research Fund, and Prof. Leung Memorial Fund, UMass Dartmouth Foundation.
\bigskip

\noindent
{\bf Appendix.

\noindent
Idealized Quark-Lepton Mass Symmetry (QLMS)}
 
   Quarks are permanently confined and do not have observable free states, in sharp contrast to other particles such as the electrons.  Thus, a quark mass may not be well-defined experimentally in the same sense as that of an electron.  It may not be accurately measured without having the help of some specific model. If the correspondence between 6 quarks and 6 leptons is not purely accidental.  Then it may be possible that there is a deeper reason behind it.  In this connection, let us explore and examine an idealized quark-lepton mass symmetry as follows.
   \bigskip  
 
 {\bf   Idealized  QLMS:}
  
  u(0.5), c(105), t(1776), \ \ \ \ \ \ \ e(0.5),  \ \ $\mu(105),  \ \  \tau(1776), $
 
$ d(\approx 0), s(\approx 0), b(\approx 0);$ \ \ \ \ \ \ \ \ \   $\nu_e(\approx 0), \nu_\mu(\approx 0), \nu_\tau(\approx 0);$
\bigskip

The idealized QLMS results for baryon spectra are compared with data as follows:  
    \be
   QLMS:  K=0.012, b= \frac{ 0.02}{\sqrt{MeV}},   \ \ \ \  (d|uu) 
  \ee
$$
  \ell=0, \   \kappa=-1;  \ \ J^P=1/2^+  \ \ \    {\bf data},  
     $$
  $$
    n=0, \ \ \  E_{0}\approx 917MeV, \ \ \ \ \ \ \ \ \ \    {\bf  N(938)}, 
  $$
  $$
   n=1,   \ \ \ E_{1}  \approx 1340 MeV, \ \ \ \ \ \ \ \ \  {\bf N(1440)}, 
  $$
  $$
     n=2,   \ \ \ E_{2} \approx 1706 MeV, \ \ \ \ \  \ \ \ \ \  {\bf N(1710)}, 
  $$
  $$
 n=3,   \ \ \ E_{3} \approx 2032 MeV, \ \ \ \ \ \ \ \ \ \ \ \   {\bf N(1880)}, 
 $$
  $$
    n=4,   \ \ \ E_{4} \approx 2333 MeV, \ \ \ \ \  \ \ \ \ \ \   {\bf N(2100)}, 
  $$
  $$
     n=5,    \ \ \ E_{5} \approx 2614 MeV, \ \ \ \ \  \ \ \ \ \ \ \   {\bf  N(2300) }, 
  $$
 
 \be
   QLMS:  K=0.012, b= \frac{ 0.038}{\sqrt{MeV}},  \ \  (d|uu)  
   \ee
$$
      \ell=1,  \   \kappa= 1;  J^P=1/2^-  \ \ \ \ \ \ \ \ \ \     {\bf data},   
      $$
   $$
    n=0, \ \  E_{0}(1/2^-)\approx 1430 MeV, \ \ \ \ \ \ \ \ \ \ \ \   {\bf N(1535)}, 
  $$
  $$
   n=1,   \ \ \ E_{1}(1/2^-)   \approx 1828 MeV, \ \ \ \ \ \ \ \ \ \   {\bf N(1650)}, 
  $$
  $$
    n=2,   \ \ \ E_{2}(1/2^-) \approx 2175 MeV, \ \ \ \ \  \ \ \ \ \  {\bf N(1895)}, 
  $$
  \be
    QLMS:  K=0.012, b= \frac{ 0.03}{\sqrt{MeV}},   \ \ \ \  (d|uu)   
  \ee
$$
    \ell=1,  \  \kappa= -2;  J^P=3/2^-  \ \ \ \ \ \     {\bf data},  
      $$
  $$ 
   n=0, \ \  E_{0}\approx 1316 MeV, \ \ \ \ \ \ \ \ \ \ \ \   {\bf  N(1520)}, 
  $$
  $$
   n=1,   \ \ \ E_{1} \approx 1717 MeV, \ \ \ \ \ \ \ \ \ \ \   {\bf N(1700)}, 
  $$
  $$
    n=2,   \ \ \ E_{2} \approx 2065 MeV, \ \ \ \ \  \ \ \ \ \ \ \   {\bf N(1875)}, 
  $$
  $$
    n=3,   \ \ \ E_{3} \approx 2380 MeV,  \ \ \ \ \  \ \ \ \ \ \ \ \  {\bf N(2120)}, 
  $$
  \be
    QLMS:  K=0.012, b= \frac{ 0.0285}{\sqrt{MeV}},   \ \ (d|uu) 
  \ee
$$
     \ \ell=2, \kappa=2; \ \ J^P=3/2^+  \ \ \ \ \ \ \ \  {\bf data},  
      $$
  $$ 
   n=0, \ \  E_{0}\approx 1487 MeV, \ \ \ \ \ \ \ \ \ \ \ \   {\bf  N(1720)}, 
  $$
  $$
   n=1,   \ \ \ E_{1} \approx 1862 MeV, \ \ \ \ \ \ \ \ \ \ \   {\bf N(1900)}, 
  $$
  $$
    n=2,   \ \ \ E_{2} \approx 2194 MeV, \ \ \ \ \  \ \ \ \ \ \ \   {\bf N(2040)}, 
  $$  
       \be
    QLMS:  K=0.012, b= \frac{ 0.0285}{\sqrt{MeV}},   \ \ \  (d|uu)   
      \ee
      $$
        \ell=3,    \kappa= 3;\  J^P=5/2^-  \ \ \ \ \ \    {\bf data}, 
      $$
   $$
      n=0, \ \  E_{0}\approx 1676 MeV, \ \ \ \ \ \ \ \ \ \ \  {\bf  N(1675)}, 
  $$
  $$
   n=1,   \ \ \ E_{1}  \approx 2028 MeV, \ \ \ \ \ \ \ \ \     {\bf N(2060)}, 
  $$
  $$
    n=2,   \ \ \ E_{2} \approx 2346 MeV, \ \ \ \ \  \ \ \ \   {\bf N(2570)}, 
  $$
$$
    n=3,   \ \ \ E_{3} \approx 2639 MeV, (prediction). 
  $$
    \be
    QLMS:  K=0.012, b= \frac{ 0.0285}{\sqrt{MeV}},   \ \ \ \  (d|uu) 
   \ee
$$
       \ell=2,  \  \kappa= -3;  J^P=5/2^+  \ \ \ \ \ \     {\bf data},   
      $$
   $$
   \ \ n=0, \ \  E_{0}\approx 1506 MeV, \ \ \ \ \ \ \ \ \ \ \ \ \ \  {\bf N(1680)}, 
  $$
  $$
 \ \ \ \ \  n=1,   \ \ \ E_{1}  \approx 1877 MeV, \ \ \ \ \ \ \ \ \ \ \ \ \  {\bf N(1860)}, 
  $$
  $$
 \ \ \ \ \   n=2,   \ \ \ E_{2} \approx 2207 MeV, \ \ \ \ \  \ \ \ \ \ \ \ \ {\bf N(2000)}, 
  $$
         \be
     QLMS:  K=0.012, b= \frac{ 0.035}{\sqrt{MeV}},  \ \ (d|uu)  
  \ee
$$
    QLMS(MeV), \ell=4, \    \kappa= 4;  \ \ \ \ \ \ \  \   {\bf data},  
      $$
   $$
  n=0, \ \  E_{0}\approx 1949 MeV, \ \ \ \  {\bf N(1990)}, 
  $$
  $$
   n=1,   \ \ \ E_{1}  \approx 2282 MeV, (prdiction). 
  $$
 \subsection{$\D$ baryons }
       \be
   QLMS:  K=0.012, b= \frac{ 0.05}{\sqrt{MeV}},  \D^+= (d|uu),   
   \ee
$$
    QLMS(MeV)    \ell=1,  \   \kappa= 1;   \ \ \ \ \ \ \ \ \      {\bf data}, 
   $$
   $$
   n=0, \ \ \ E_0\approx 1575 MeV, \ \ \ \ \ \ \ \ \ \ \ \ \ \   {\bf \D(1620)}, 
   $$
   $$
   n=1, \ \ \  E_1 \approx 1967 MeV, \ \ \ \ \ \ \ \ \ \ \ \ \ \ \   {\bf \D(1900)}, 
   $$
   $$  
     n=2,   \ \ \ E_{2} \approx 2310 MeV, \ \ \ \ \  \ \ \ \ \ \ \ \ \ \ \ \  {\bf \D(2150)},  
  $$
  \be
   QLMS:  K=0.012, b= \frac{ 0.067}{\sqrt{MeV}},    \D^+= (d|uu),  \ \ \ \ \ 
   \ee
$$
   QLMS(MeV), \ell=0,  \   \kappa= -1;  \ \ \ \ \ \ \ \ \ \ \ \ \     {\bf data},  
      $$
   $$
  n=0, \ \  E_{0} \approx 1711 MeV, \ \ \ \ \ \ \ \ \ \ \ \ \ \ \ \ \ \   {\bf \D(1750)}, 
  $$
  $$
  n=1,   \ \ \ E_{1}  \approx 2115 MeV, \ \ \ \ \ \ \ \ \ \ \ \ \ \ \ \ \ \  {\bf \D(1910)},  
  $$
  $$
 n=2,   \ \ \ E_{2} \approx 2466 MeV, (prediction). \ \ \ \ \ 
$$
     \be
   QLMS:  K=0.012, b= \frac{ 0.049}{\sqrt{MeV}},  \D^+= (d|uu), 
  \ee
$$
  QLMS(MeV),  \ell=1,   \  \kappa= -2; \ \ \ \ \ \         {\bf data},  
      $$
   $$
   n=0, \ \  E_{0} \approx 1622 MeV, \ \ \ \ \ \ \ \ \ \ \ \ \ \ \ \ \ \ \   {\bf  \D(1700)}, 
  $$
  $$
   n=1,   \ \ \ E_{1}  \approx 2009 MeV, \ \ \ \ \ \ \ \ \ \ \ \ \ \ \ \ \  {\bf \D(1940)}, 
  $$
 $$
   n=2,   \ \ \ E_{2} \approx 2350 MeV, (prediction). \ \ \ \ \ \ \ \ \ \ \ \ \ \ \ \ \ \ \ \  
  $$
    
 \subsection{$\L$ baryons}
 For $\L^{0}(uds)$ baryon states with zero charge and even parity, i.e., $N1/2^+$,the confining QCD model gives the following results for $E_n\approx E_s(n) + E_{ud}(n)$ with the corresponding experimental data.\cite{3}  

\be
 QLMS : K = 0.006,b = \frac{0.056}{\sqrt{MeV}}, \ \ \ \     \L^0=(s|ud),
    \ee
  $$
 \ell=0, \ \ \kappa=-1;  J^P=1/2^+ \ \ \ \ \ \ \ \ \ {\bf data}, 
$$
  $$
   n=0, \ \  E_{0} \approx 1108 MeV, \ \ \ \ \ \ \ \ \ \ \ \  {\bf  \L(1115)}, 
  $$
  $$
  n=1,   \ \ \ E_{1}  \approx 1443 MeV, \ \ \ \ \ \ \ \ \ \ \ \ \  {\bf \L(1600)}, 
   $$
  $$
   n=2,   \ \ \ E_{2} \approx 1728 MeV, \ \ \ \ \  \ \ \ \ \ \ \ \   {\bf \L(1710)}, 
  $$
     $$
   n=3,    \ \ \ E_{3} \approx 1984 MeV, \ \ \ \ \  \ \ \ \ \ \ \   {\bf \L(1810)}, 
  $$
      $$
   n=4,    \ \ \ E_{4} \approx 2219 MeV, (prediction). \ \ \ \ \ 
  $$

  \be
  QLMS : K = 0.006,b = \frac{0.056}{\sqrt{MeV}},  \L^0=(s|ud), \ \ \ \  
  \ee
  $$
   \ell=1,   \   \kappa= 1;  \ \ \ \ \  \ {\bf data}, 
$$
  $$
   n=0, \ \  E_{0}\approx 1281 MeV, \ \ \ \ \ \ \ \ \ \ \ \ \ \ \ \  {\bf  \L(1380)}, 
  $$
  $$
   n=1,   \ \ \ E_{1}  \approx 1588 MeV, \ \ \ \ \ \ \ \ \ \ \ \ \ \ \ \ \  {\bf \L(1405)}, 
  $$
  $$
    n=2,   \ \ \ E_{2} \approx 1858 MeV,  \ \ \ \ \ \ \ \ \ \ \ \ \ \ \ \ \  {\bf \L(1670)}, 
  $$
     $$
   n=3,    \ \ \ E_{3} \approx 2103 MeV,  \ \ \ \ \  \ \ \ \ \ \ \ \ \ \ \ \   {\bf \L(1800)}, 
  $$
       $$
  n=4,    \ \ \ E_{4} \approx 2331 MeV,   \ \ \ \  \ \ \ \ \ \ \ \ \ \ \ \ \ \ \   {\bf \L(2000)}, 
  $$
  
    \subsection{$\S$ baryons}okokok
     \be
   QLMS : K = 0.006,b = \frac{0.07}{\sqrt{MeV}},  \S^+=(s|uu),  
  \ee
  $$
 \ell=0, \  \kappa= -1;  \ \  J^P=1/2^+\ \ \ \ \ \ \ \  {\bf data}, okok
$$
  $$
   n=0, \ \  E_{0} \approx 1284 MeV, \ \ \ \ \ \ \ \ \ \ \ \ \ \ \  {\bf  \S(1189)}, 
  $$
  $$
  n=1,   \ \ E_{1}  \approx 1606 MeV, \ \ \ \ \ \ \ \ \ \ \ \ \ \ \ \ \  {\bf \S(1660)}, 
   $$
  $$
    n=2,   \ \ \ E_{2} \approx 1885 MeV, \ \ \ \ \  \ \ \ \ \ \ \ \ \ \ \ \  {\bf \S(1880)}, 
  $$
      \be
  QLMS : K = 0.006,b = \frac{0.073}{\sqrt{MeV}},   \S^+=(s|uu), 
  \ee
  $$
  \ell=1, \  \kappa= 1; \ \ J^P=1/2^- \ \ \ \ \ \ \ \ \ \ \ \  {\bf data}, 
$$
   $$
   n=0, \ \  E_{0} \approx 1484 MeV, \ \ \ \ \ \ \ \ \ \ \ \   {\bf \S(1620)}, 
  $$
  $$
   n=1,   \ \ \ E_{1}   \approx 1781 MeV, \ \ \ \ \ \ \ \ \ \ \ \ \  {\bf \S(1750)}, 
  $$
   $$
   n=2,   \ \ \ E_{2}  \approx 2045 MeV, \ \ \ \ \ \ \ \ \ \ \ \ \  {\bf \S(1900)}, 
  $$
  $$
    n=3,   \ \ \ E_{3} \approx 2286 MeV,  \ \ \ \ \ \ \ \ \ \ \ \ \ \   {\bf \S(2160)},  
  $$
     \be
  QLMS : K = 0.006,b = \frac{0.05}{\sqrt{MeV}},   \S^+=(s|uu), 
 \ee
  $$
   \ell=2, \  \kappa= 2;  \ \ J^P=3/2^+ \ \ \ \ \ \ \ \ \    {\bf data}, 
$$
 $$
    n=0, \ \  E_{0} \approx 1370 MeV, \ \ \ \ \ \ \ \  \  {\bf  \S(1385)}, 
  $$
  $$
  n=1,   \ \ \ E_{1}  \approx 1659 MeV, \ \ \ \ \ \ \ \    {\bf  \S(1780)},
  $$
   $$
  n=2,   \ \ \ E_{2} \approx 1917MeV, \ \ \ \ \ \ \ \   {\bf \S(1940)}, 
  $$
   $$
  n=3,   \ \ \ E_{3}  \approx 2152 MeV, \ \ \ \ \ \ \ \ \ \  {\bf \S(2080)}, 
  $$
   $$
  n=4,   \ \ \ E_{4} \approx 2376 MeV, \ \ \ \ \ \ \ \    {\bf \S(2230)}, 
  $$
 $$
    n=5,   \ \ \ E_{5} \approx 2585 MeV, (prediction). okok\ \ \ \ \ \ 
  $$ 
  \be
  QLMS : K = 0.012,  b = \frac{0.033}{\sqrt{MeV}}, \S^+=(s|uu), 
  \ee
  $$
  \ell=1, \  \kappa= -2;  \ \ J^P=3/2^-\ \ \ \ \ \ \ \ \ \ \ \   {\bf data}, 
$$
   $$
    n=0,   \ \ \ E_{0}  \approx 1428 MeV, \ \ \ \ \  \ \ \ \ \ \ \ \  {\bf \S(1580)}, 
  $$
  $$
    n=1,   \ \ \ E_{2} \approx 1832 MeV, \ \ \ \ \  \ \ \ \ \ \ \ \   {\bf \S(1670)}, 
  $$
   $$
   n=2,   \ \ \ E_{3}\approx 2184 MeV, \ \ \ \ \  \ \ \ \ \ \ \ \  {\bf \S(1910)}, 
  $$
   $$
     n=3,   \ \ \ E_{4}\approx 2503 MeV, \ \ \ \ \  \ \ \ \ \ \ \ \  {\bf \S(2010)}, 
  $$

  \subsection{{\bf Omega} and charmed  baryons}
 \bigskip
\noindent
{\bf Omega baryons}  
   \be
  QLMS : K = 0.006, \ \ b = \frac{0.096}{\sqrt{MeV}}, \Om^-=(s|ss), \ \  \ \ \ \ 
  \ee
  $$
  \ell=0, \  \kappa= -1;   J^P=1/2^+ \ \ \ \ \ \ \ \ \ \    {\bf data},  
$$
  $$
   n=0, \ \  E_{0} \approx 1614 MeV, \ \ \ \ \ \ \ \ \ \ \ \ \ \ \  {\bf  \Om(1672)}, 
  $$
  $$
  n=1,   \ \ E_{1}  \approx 1917 MeV, \ \ \ \ \ \ \ \ \ \ \ \ \ \ \    {\bf \Om(2012)}, 
   $$
    $$
    n=2,   \ \ \ E_{2} \approx 2184 MeV,   \ \ \ \ \ \ \ \ \ \ \ \ \ \ \   {\bf \Om(2250)}, 
  $$
  $$
  n=3,   \ \ E_{3}  \approx 2429 MeV, \ \ \ \ \ \ \ \ \ \ \ \ \ \ \    {\bf \Om(2470)}, 
   $$
  $$
    n=4,   \ \ \ E_{2} \approx 2657 MeV, (prediction).  \ \ \ \ \ \ \ \ \ \ \ \ 
  $$
\bigskip
\noindent
{\bf Charmed baryons}
\be
    QLMS : K = 0.006, \ \ b = \frac{0.14}{\sqrt{MeV}}, \L^+_c=(c|ud), \ \ \ \ \ 
  \ee
  $$
  \ell=0, \  \kappa= -1;  J^P=1/2^+,   \ \ \ \ \ \ {\bf data},  
$$
  $$
   n=0, \ \  E_{0} \approx 2286 MeV, OK    \ \ \ \ \ \ \ \ \ \ \ \ \  {\bf  \L_c(2286)}, 
  $$
  $$
    n=1,   \ \ \ E_{1} \approx 2551 MeV, (prediction). \ \ \ \ \  \ \    
  $$
\bigskip
 \be
    QLMS : K = 0.006, \ \ b = \frac{0.16}{\sqrt{MeV}}, \L_c=(c|ud), \ \   \ \ \ \ 
  \ee
  $$
  \ell=1, \  \kappa= -2; \  J^P=3/2^-   \ \ \ \ \ \ \ \ \ \  {\bf data},  
$$
  $$
   n=0, \ \  E_{0} \approx 2686 MeV, \ \ \ \ \ \ \ \ \ \ \ \ \ \ \  {\bf  \L_c(2625)}, 
  $$
   $$
   n=1, \ \  E_{1} \approx 2930 MeV, \ \ \ \ \ \ \ \ \ \ \ \ \ \ \  {\bf  \L_c(2940)}, 
  $$
  $$
    n=2,   \ \ \ E_{2} \approx 3157 MeV, (prediction). \ \ \ \ \  \ \ \ \ \ \ \ \ \ \ \ \  
  $$ 
\noindent
\subsection{{\bf Bottom baryons}}
\be
  QLMS : K = 0.006, \ \ b = \frac{0.39}{\sqrt{MeV}},  \ \  \Ld^+_b=(b|ud), 
 \ee
  $$
   \ell=0, \  \kappa= -1;  \ \  J^P=1/2^+ \ \ \ \ \ \     {\bf data},  
$$
  $$
   n=0, \ \  E_{0} \approx 5697 MeV, \ \ \ \ \ \ \    {\bf  \Ld_b^0(5620)},   
  $$
  $$
    n=1,   \ \ \ E_{1} \approx 5880 MeV, (prediction)  \ \ \ \ \  \ \ \  
  $$
  \be
   QLMS : K = 0.006, \ \ b = \frac{0.41}{\sqrt{MeV}},   \ \     \Xi_b=(b|ds), \ \ \ \ 
  \ee
  $$
   \ell=0, \  \kappa= -1;   J^P=1/2^+ \ \ \ \ \ \ \      {\bf data},  
$$
  $$
   n=0, \ \  E_{0} \approx 5871 MeV, \ \ \ \ \ \ \ \ \   {\bf  \Xi^0_b(5800)}, 
  $$
  $$
   n=1, \ \  E_{1} \approx 6052 MeV, \ \ \ \ \ \ \ \ \   {\bf  \Xi'_b(5935)}, 
  $$
  $$
    n=2,   \ \ \ E_{2} \approx 6228 MeV, (prediction). \ \ \ \ \  \ \ \  
  $$
  \bigskip
   In general, the agreement for the idealized quark-lepton mass symmetry has the same approximations as those in (28)-(90) in the previous work.\cite{1}  The reason is that the eigenvalues for the baryon mass spectra appear to be very insensitive to the quark masses in the confining quarkdynamics model.

 \bibliographystyle{unsrt}



\end{document}


\clearpage

\end{document}

 about the same..????? below
  \be
  model:  \ell=0, \   \kappa=-1; \ \ \ \ \    \  p, N^+ = uud, \ \  (n, N^0=udd)
  \ee
    $$
    CQD(I):  K=0.012, b= \frac{ 0.014}{\sqrt{MeV}},   \ \ \ \ \ \ \ \ \      \left[CQD(II) :  same \ K, b  \right],   
   $$
$$
     CQD(I),  \ \ \ \ \ \ \ \ \ \ \ \ \ \ \ \ \ \ \ \ \ \ \ \ \     {\bf data},   \ \ \ \ \ \ \ \ \    [CQD(II)],      
      $$
  $$
    n=0, \ \  E_{0}\approx 908MeV, \ \ \ \ \ \ \ \ \ \   {\bf  N(938.3)}, \ \ \ \ [908], \ \  
  $$
  $$
   n=1,   \ \ \ E_{1}  \approx 1342 MeV, \ \ \ \ \ \ \ \ \  {\bf N(1440)}, \ \ \ \ [1340], \ \  
     $$
  $$
     n=2,   \ \ \ E_{2} \approx 1707 MeV, \ \ \ \ \  \ \ \ \ \  {\bf N(1710)}, \ \ \ \ [1700], \ \ \  
  $$
  $$
 n=3,   \ \ \ E_{3} \approx 2032 MeV, \ \ \ \ \ \ \ \ \ \ \ \   {\bf N(1880)}, \ \ \ \ [2034], \ \ \  
  $$
  $$
    n=4,   \ \ \ E_{4} \approx 2335 MeV, \ \ \ \ \  \ \ \ \ \ \   {\bf N(2100)}, \ \ \ \ [2333], \ \ \  
  $$
   $$
     n=5,    \ \ \ E_{5} \approx 2615 MeV, \ \ \ \ \  \ \ \ \ \   {\bf  N(2300) }, \ \ \ \  [2615], \ \ \  
  $$
   $$
    n=6,    \ \ \ E_{6} \approx 2881 MeV, (prediction)  \ \ \ \  \ \ \ \ \ \ \     [2881].
  $$
    \bigskip
    
  \be
   model:   \  \ell=1,  \   \kappa= 1; \ \ \ \ \   \      p, N^+ = uud, \ \  (n, N^0=udd)
  \ee 
    $$
    CQD(I):  K=0.012, b= \frac{ 0.028}{\sqrt{MeV}},   \ \ \ \ \ \ \ \ \     \left[CQD(II):   same \ K, b   \right],  
   $$
$$
    CQD(I),  \ \ \ \ \ \ \ \ \ \ \ \ \ \ \ \ \ \ \ \ \ \ \ \ \ \ \ \ \ \ \ \ \ \  {\bf data},   \ \ \ \ \ \ \ \ \  \ \ \    [CQD(II)],             
    $$
   $$
    n=0, \ \  E_{0}\approx 1387 MeV, \ \ \ \ \ \ \ \ \ \ \ \ \ \ \ \ \ \ \ \ \  {\bf N(1535)}, \ \ \ \ \ \ \ [1387], \ \ 
  $$
  $$
   n=1,   \ \ \ E_{1}(1/2^-)   \approx 1766 MeV, \ \ \ \ \ \ \ \ \ \ \   {\bf N(1650)}, \ \ \ \ \ \ \ \ [1766], \  
  $$
  $$
    n=2,   \ \ \ E_{2}(1/2^-) \approx 2101 MeV, \ \ \ \ \  \ \ \ \ \  {\bf N(1895)}, \ \ \ \ \ \ \ [2101], \ \ 
  $$
  $$
    n=3,   \ \ \ E_{3}(1/2^-) \approx 2408 MeV, (prediction). \ \ \ \ \  \ \ \ \ \ \ \ \ \  [2408]. 
   $$

   \bigskip
   
  \be
   model: \   \ell=1,  \  \kappa= -2; \ \ \ \ \ \  \      p, N^+ = uud, \ \  (n, N^0=udd)\ 
  \ee 
    $$
    CQD(I):  K=0.012, b= \frac{ 0.026}{\sqrt{MeV}},   \ \ \ \ \ \ \ \ \ \ \        \left[CQD(II):    same \ K, b  \right],  
   $$
$$
    CQD(I),  \ \ \ \ \ \ \ \ \ \ \ \ \ \ \ \ \ \ \ \ \ \ \ \ \ \ \ \ \ \ \   {\bf data},   \ \ \ \ \ \ \ \ \ \ \ \ \ \    [CQD(II)],     
      $$
  $$ 
   n=0, \ \  E_{0}\approx 1362 MeV, \ \ \ \ \ \ \ \ \ \ \ \   {\bf  N(1520)}, \ \ \ \ \ \ \ [1362], \ 
  $$
  $$
   n=1,   \ \ \ E_{1} \approx 1741 MeV, \ \ \ \ \ \ \ \ \ \ \   {\bf N(1700)}, \ \ \ \ [1741], \  
  $$
  $$
    n=2,   \ \ \ E_{2} \approx 2076 MeV, \ \ \ \ \  \ \ \ \ \ \ \   {\bf N(1875)}, \ \ \ \ \ [2076], \  
  $$
  $$
    n=3,   \ \ \ E_{3} \approx 2383 MeV,  \ \ \ \ \  \ \ \ \ \ \ \ \  {\bf N(2120)}, \ \ \ \ [2383], \ \ 
  $$
$$
   n=4,   \ \ \ E_{4} \approx 2669 MeV, (prediction). \ \ \ \ \ \ \ \ \ \ \ \ \  \   [2669]. \ \ 
     $$
      \bigskip
  \be
  model: \ \ell=2, \kappa=2; \ \ \ \ \ \ \   p, N^+ = uud, \ \  (n, N^0=udd)
  \ee
   $$
    CQD(I):  K=0.012, b= \frac{ 0.026}{\sqrt{MeV}},   \ \ \ \ \ \ \ \ \ \    \left[CQD(II):   same \ K, b  \right],  
   $$
$$
    CQD(I),  \ \ \ \ \ \ \ \ \ \ \ \  \ \ \ \ \ \ \ \ \ \ \ \ \ \ \ \ \ \ \ \ \ \ \  {\bf data},   \ \ \ \ \ \ \ \ \ \ \ \ \ \    [CQD(II)],      
      $$
  $$ 
   n=0, \ \  E_{0}\approx 1579 MeV, \ \ \ \ \ \ \ \ \    {\bf  N(1720)}, \ \ \ \ [1549], \  
     $$
  $$
   n=1,   \ \ \ E_{1} \approx 1935 MeV, \ \ \ \ \ \ \ \ \ \ \   {\bf N(1900)}, \ \ \ \ [1905], \ \ 
  $$
  $$
    n=2,   \ \ \ E_{2} \approx 2255 MeV, \ \ \ \ \  \ \ \ \ \ \ \   {\bf N(2040)}, \ \ \ \ \ [2226], \ \ 
  $$  
    $$
   n=3,   \ \ \ E_{3} \approx 2551 MeV, (prediction). \ \ \ \ \ \ \ \ \ \ \ \ \  \   [2522].\ \ 
  $$
  \bigskip
  
       \be
   model:   \ell=3,    \kappa= 3; \ \ \ \ \   \ \  p, N^+ = uud, \ \  (n, N^0=udd)   
   \ee
     $$
    CQD(I):  K=0.012, b= \frac{ 0.025}{\sqrt{MeV}},   \ \ \ \ \ \ \ \ \ \ \      \left[CQD(II):   same \ K, b    \right],     
      $$
      $$
     CQD(I), \ \ \ \ \ \ \ \ \ \ \ \ \ \ \ \ \ \ \ \ \ \ \ \ \ \ \ \ \ \ \ \ \ {\bf data}, \ \ \ \ \ \ \ \ \ \  [CQD(II)], \ 
      $$
   $$
      n=0, \ \  E_{0}\approx 1705 MeV, \ \ \ \ \ \ \ \ \ \ \  {\bf  N(1675)},  \ \ \ \ [1706], \ 
  $$
  $$
   n=1,   \ \ \ E_{1}  \approx 2044 MeV, \ \ \ \ \ \ \ \ \     {\bf N(2060)}, \ \ \ \ (2045), \  
     $$
  $$
    n=2,   \ \ \ E_{2} \approx 2354 MeV, \ \ \ \ \  \ \ \ \   {\bf N(2570)}, \ \ \ \ \ \ [2353], \  
  $$
$$
    n=3,   \ \ \ E_{3} \approx 2641 MeV, (prediction). \ \ \ \ \ \ \ \ \ \ \ \ \ \  [2641].  
  $$
      \bigskip
      
  \be
   model: \   \ell=2,  \  \kappa= -3; \ \ \ \ \   \    p, N^+ = uud, \ \  (n, N^0=udd)
  \ee 
    $$
    CQD(I):  K=0.012, b= \frac{ 0.025}{\sqrt{MeV}},   \ \ \ \ \ \ \ \ \ \ \ \       \left[(CQD(II):   same \ K, b  \right],  
   $$
$$
     CQD(I),  \ \ \ \ \ \ \ \ \ \ \ \ \ \ \ \  \ \ \ \ \ \ \ \ \ \ \ \ \ \ \ \  {\bf data},   \ \ \ \ \ \ \ \ \ \ \     [ CQD(II)],       
      $$
   $$
   \ \ n=0, \ \  E_{0}\approx 1544 MeV, \ \ \ \ \ \ \ \ \ \ \ \ \ \  {\bf N(1680)}, \ \ \ \ \ [1543], \  
  $$
  $$
 \ \ \ \ \  n=1,   \ \ \ E_{1}  \approx 1899 MeV, \ \ \ \ \ \ \ \ \ \ \ \ \  {\bf N(1860)}, \ \ \ \ \ [1898], \ 
  $$
  $$
 \ \ \ \ \   n=2,   \ \ \ E_{2} \approx 2219 MeV, \ \ \ \ \  \ \ \ \ \ \ \ \ {\bf N(2000)}, \ \ \ \ \ [2222], \  
  $$
$$
 \ \ \ \ \   n=3,   \ \ \ E_{3} \approx 2514 MeV, (prediction). \ \ \ \ \  \ \ \ \ \ \ \ \ \  [2514].  
  $$

  \be
   model:   \  \ell=4, \    \kappa= 4; \ \ \ \  \    p, N^+ = uud, \ \  (n, N^0=udd)
 \ee  
   $$
    CQD(I):  K=0.012, b= \frac{ 0.025}{\sqrt{MeV}},   \ \ \ \      \ \ \ \ \ \ \ \       \left[(CQD(II):   same \ K, b  \right],   
   $$
$$
    CQD(I),  \ \ \ \ \ \ \ \ \ \ \ \ \ \ \ \  \ \ \ \ \ \ \ \ \ \ \ \ \ \ \ \  {\bf data},   \ \ \ \ \ \ \ \ \ \ \     [ CQD(II)],        
      $$
   $$
  n=0, \ \  E_{0}\approx 1893 MeV, \ \ \ \  {\bf N(1990)}, \ \ \ \ \ \ \ \ \ [1851],  
  $$
  $$
   n=1,   \ \ \ E_{1}  \approx 2214 MeV, (prdiction). \ \ \ \ \ \ \ \ \ \ \ \ \  [2167].
  $$
     \bigskip
     
  \be
   model:  \   \ell=3, \  \kappa= -4; \ \ \ \ \      p, N^+ = uud, \ \  (n, N^0=udd)
   \ee
     $$
    CQD(I):  K=0.012, b= \frac{ 0.05}{\sqrt{MeV}},   \ \ \ \  \ \ \ \ \ \ \ \       \left[(CQD(II):   same \ K, b  \right],  
   $$
$$
      CQD(I),  \ \ \ \ \ \ \ \ \ \ \ \ \ \ \ \  \ \ \ \ \ \ \ \ \ \ \ \ \ \ \ \  {\bf data},   \ \ \ \ \ \ \ \ \ \ \     [ CQD(II)],       
      $$
   $$
   n=0, \ \  E_{0}\approx 2169 MeV, \ \ \ \ \ \ \ \ \ \  {\bf N(2190)}, \ \ \ \ \ \ \ \ \ \ \ [2122],  
  $$
  $$
    n=1,   \ \ \ E_{1}   \approx 2488 MeV, (prediction), \ \ \ \ \ \ \ \ \ \ \ \ \ \ \ \ \ \  [2436].
  $$

      \bigskip
      
  \be
   model:  \  \ell=4, \   \kappa= -5; \ \ \ \     \   p, N^+ = uud, \ \  (n, N^0=udd)
      \ee
     $$
    CQD(I):  K=0.012, b= \frac{ 0.04}{\sqrt{MeV}},   \ \ \ \    \ \ \ \ \ \ \ \       \left[(CQD(II):   same \ K, b  \right],   
   $$
$$
      CQD(I),  \ \ \ \ \ \ \ \ \ \ \ \ \ \ \ \  \ \ \ \ \ \ \ \ \ \ \ \ \ \ \ \  {\bf data},   \ \ \ \ \ \ \ \ \ \ \     [ CQD(II)],          
      $$
   $$
   n=0, \ \  E_{0} \approx 2170 MeV, \ \ \ \ \ \ \ \ \  {\bf N(2220)}, \ \ \ \ \ [2108],  
  $$
  $$
  n=1,   \ \ \ E_{1}   \approx 2479 MeV,  (prediction). \ \ \ \ \ \ \ \ \ \ \  [2416].
  $$
      \bigskip
      
  \be
   model:  \  \ell=5,  \  \kappa= 5; \ \ \ \ \   \   p, N^+ = uud, \ \  (n, N^0=udd)
   \ee
     $$
    CQD(I):  K=0.012, b= \frac{ 0.035}{\sqrt{MeV}},   \ \ \ \   \ \ \ \ \ \ \ \       \left[(CQD(II):   same \ K, b  \right],  
   $$
$$
      CQD(I),  \ \ \ \ \ \ \ \ \ \ \ \ \ \ \ \  \ \ \ \ \ \ \ \ \ \ \ \ \ \ \ \  {\bf data},   \ \ \ \ \ \ \ \ \ \ \     [ CQD(II)],           
      $$
   $$
    n=0, \ \  E_{0})\approx 2227 MeV, \ \ \ \ \ \ \ \ \ \  {\bf  N(2250)},\ \ \ \ \ \ \ \ \ \ \ \ \ \  [2203], 
  $$
  $$
    n=1,   \ \ \ E_{1}   \approx 2529 MeV, (prediction). \ \ \ \ \ \ \ \ \ \ \ \ \ \ \ \ \ \ \ \ \ \ [2510].
  $$
  \bigskip

  \be
   model:  \   \ell=5, \   \kappa= -6; \ \ \ \ \  \     p, N^+ = uud, \ \  (n, N^0=udd)
   \ee
       $$
    CQD(I):  K=0.012, b= \frac{ 0.06}{\sqrt{MeV}},   \ \ \ \   \ \ \ \ \ \ \ \       \left[(CQD(II):   same \ K, b  \right],  
   $$
$$
      CQD(I),  \ \ \ \ \ \ \ \ \ \ \ \ \ \ \ \  \ \ \ \ \ \ \ \ \ \ \ \ \ \ \ \  {\bf data},   \ \ \ \ \ \ \ \ \ \ \     [ CQD(II)],         
      $$
   $$
   n=0, \ \  E_{0}\approx 2673 MeV, \ \ \ \ \ \ \ \ \ \  {\bf N(2600)}, \ \ \ \ \ \ \ \ \ \ \ \ \ \  [2665], 
  $$
  $$
   n=1,   \ \ \ E_{1}   \approx 2962 MeV, (prediction). \ \ \ \ \ \ \ \ \ \ \ \ \ \ \ \ \ \ \ \ \   [2956].  
  $$
 \bigskip
  \be
   model: \  \ell=6,  \  \kappa= -7; \ \ \ \ \ \    p, N^+ = uud, \ \  (n, N^0=udd)
 \ee  
     $$
    CQD(I):  K=0.012, b= \frac{ 0.05}{\sqrt{MeV}},   \ \ \   \ \ \ \ \ \ \ \       \left[(CQD(II):   same \ K, b  \right],  
   $$
$$
      CQD(I),  \ \ \ \ \ \ \ \ \ \ \ \ \ \ \ \  \ \ \ \ \ \ \ \ \ \ \ \ \ \ \ \  {\bf data},   \ \ \ \ \ \ \ \ \ \ \     [ CQD(II)],       
       $$
   $$
 n=0, \ \  E_{0}\approx 2650 MeV, \ \ \ \  {\bf  N(2700)}, \ \ \ \ \ \ \ \ \ \ \ \ \ \ \ \  [2643], 
  $$
  $$
   n=1,   \ \ \ E_{1}   \approx 2933 MeV, (prediction). \ \ \ \ \ \ \ \ \ \ \ \ \ \ \ \  [2930].
     $$